\documentclass[aps,pre,twocolumn,groupedaddress,superscriptaddress,showpacs]{revtex4-1}

\usepackage{amsfonts,amssymb,amsbsy,amsmath}
\usepackage[pdftex]{graphicx}
\usepackage{epstopdf}
\usepackage{amsthm}
\usepackage{subfigure}
\usepackage{hhline}
\usepackage{soul}

\usepackage{color}

\newcommand{\avrg}[1]{\left\langle #1 \right\rangle}
\newcommand{\nn}{\nonumber \\}

\begin{document}

\title{Hierarchical mutual information for the comparison of hierarchical community structures in complex networks}

\author{Juan Ignacio Perotti}
\affiliation{IMT Institute for Advanced Studies Lucca, Piazza San Francesco 19, I-55100, Lucca, Italy}
\email[E-mail: ]{juanignacio.perotti@imtlucca.it}

\author{Claudio Juan Tessone}
\affiliation{URPP Social Networks, Universit\"at Z\"urich, Andreasstrasse 15, CH-8050 Z\"urich, Switzerland}
\email[E-mail: ]{claudio.tessone@business.uzh.ch}

\author{Guido Caldarelli} 
\affiliation{IMT Institute for Advanced Studies Lucca, Piazza San Francesco 19, I-55100, Lucca, Italy}
\affiliation{Institute for Complex Systems CNR, via dei Taurini 19, I-00185, Roma, Italy}
\affiliation{London Institute for Mathematical Sciences, 35a South St. Mayfair, London W1K 2XF UK}
\email[E-mail: ]{guido.caldarelli@imtlucca.it}

\date{\today}

\begin{abstract}
The quest for a quantitative characterization of community and modular structure of complex networks produced a variety of methods and algorithms to classify different networks. 
However, it is not clear if such methods provide consistent, robust and meaningful results when considering hierarchies as a whole.
Part of the problem is the lack of a similarity measure for the comparison of hierarchical community structures.
In this work we give a contribution by introducing	 the {\it hierarchical mutual information}, which is a generalization of the traditional mutual information, and allows to compare hierarchical partitions and hierarchical community structures. 
The {\it normalized} version of the hierarchical mutual information should behave analogously to the traditional normalized mutual information.
Here, the correct behavior of the hierarchical mutual information is corroborated on an extensive battery of numerical experiments.
The experiments are performed on artificial hierarchies, and on the hierarchical community structure of artificial and empirical networks.
Furthermore, the experiments illustrate some of the practical applications of the hierarchical mutual information.
Namely, the comparison of different community detection methods, and the study of the consistency, robustness and temporal evolution of the hierarchical modular structure of networks. 
\end{abstract}

\maketitle


\section{Introduction} 

\label{sec:intro}

Many complex systems exhibit some degree of organization at different physical scales.
Often, the organization is hierarchical.
There exist examples of this fact in variegated fields, like biological, social and technological systems.  
Among the former, and starting from complex molecules (such as lipids, proteins,  RNA or DNA)  while increasing the scale of observation, new levels of organization are found: organelles, cells,  tissues, organs, anatomical systems, organisms, populations and ecosystems. 
In the social context, human societies organize from the level of individuals, groups, cities, up to the global scale of countries or continents.
Finally, among technological systems, computer networks are also arranged at different scales from the local network level up to the domain level routing systems that constitute the backbone of internet.
Hierarchical organizations seem ubiquitous in complex systems and, despite the early interest of the scientific community about the subject~\cite{simon1962architecture,anderson1972more,holland1998emergence}, it is far from being fully understood.
The description of hierarchical organization of complex systems remains, to a great extent, at the semantic level.
This is mainly because the following difficulties: the existence of several relevant physical scales,
the existence of a variety of organizing principles, the large number of components, and the lack of a generally enough and well defined formal theory for the identification of hierarchies.

The study of complex networks~\cite{boccaletti2006complex,caldarelli2007scale,newman2010networks,kivela2014multilayer} plays a central role in the characterization of the organization of complex systems.
In essence, networks are used to represent the structure of the interactions between the components of the system under consideration.
Therefore, it is reasonable to assume that some complex networks have hierarchically organized topologies, reflecting the underlying hierarchical organization of the associated complex systems.
A natural way of thinking about hierarchical network topologies is that of hierarchical community structures; i.e.~communities within communities of nodes~\cite{rosvall2011multilevel,peixoto2014hierarchical,macmahon2015community}.
Typically, the identification of the communities of a network is computationally intensive and a statistically difficult problem~\cite{fortunato2010community}.
Although a large number of community detection methods have been developed already~\cite{girvan2002community,newman2006finding,heimo2008detecting,lancichinetti2009community,zlatic2010topologically,zhang2014scalable} -- including methods for the identification of hierarchical community 
structures~
\cite{sales2007extracting,
clauset2008hierarchical,arenas2008analysis,
lancichinetti2009detecting,lancichinetti2011finding,
rosvall2011multilevel,granell2012hierarchical,
granell2012unsupervised,peixoto2014hierarchical,
queyroi2014assessing,zhang2014scalable} -- not all  methods provide comparable results.
This is true, specially for hierarchical community structures.
Therefore, similarity measures for the comparison of hierarchical community structures are of crucial importance.
The aim of this paper is to introduce an information-theoretic tool which can be used to compare hierarchies, or trees, which might be composed of network communities.
We further show that this tool can be employed to trace the evolution of hierarchies when temporal networks are analyzed.

A standard way to quantify the similarity of two community structures is to compute the mutual information between the associated node partitions~\cite{danon2005comparing,meilua2007comparing}.
Extending the idea, the present paper introduces a {\it hierarchical mutual information}, generalizing the traditional mutual information to work with hierarchical partitions.
In principle, there might be different ways in which the mutual information can be generalized into a hierarchical mutual information.
In this work, hierarchies are considered to be of divisive nature; i.e.~the whole is divided into parts, each of which is sub-divided into sub-parts, an so on, following a top-down approach.
As a consequence, in this context hierarchies are represented by trees with branches of varying length.
Other possible generalization approaches might exist. For example, generalizations that consider agglomerative hierarchies -- i.e. bottom up approaches -- or overlapping communities.
Alternatively, related methods exists for the comparison of phylogenetic trees~\cite{dasgupta1997distances,nielsen2011sub,shi2013distances}.
Recently, a method to compare hierarchies was introduced; the method follows a combinatorial approach~\cite{queyroi2015suppression}.
However, to the best of our knowledge, no previous method based on information-theoretic measures, exists for the comparison of hierarchies.
These alternative methods, and the previously mentioned  generalization approaches, are not discussed further in this paper, but can be considered in future works.

The outline of the paper is the following.
In section~\ref{sec:theo}, the hierarchical mutual information is motivated and introduced.
In section~\ref{sec:resu}, this measure is tested on different synthetic setups.
More specifically, in sub-section~\ref{sec:resu:arthier},
the behavior of the hierarchical mutual information is tested in artificial hierarchies;
while in sub-section~\ref{sec:resu:artnets}, the hierarchical mutual information is used to analyze the hierarchical community structure of artificial networks, or network models.
A similar procedure is performed on empirical networks in section~\ref{sec:resu:empnets}, including the case of a temporal one.
Finally, the discussion and conclusions are summarized in section~\ref{sec:conclu}.

\section{Theory}
\label{sec:theo}

\subsection{Hierarchical Partitions}

A hierarchical partition is a generalization of the traditional concept of partition. Here, each element of the partition can be recursively partitioned into others, yielding a hierarchy.
The formal definition is as follows.
Consider a set of elements, or universe, denoted by $\Omega$.
An element in $\Omega$ is denoted by $i$.
The set $\Omega$ splits into a hierarchy of sub-sets, denoted by $v$.
The number of elements in the sub-set $v$ is written as $|v|$.
The  {\it hierarchical partition}, or simply hierarchy, is represented by a tree denoted by $\mathcal{T}$.
The root $v_{\Omega}\in\mathcal{T}$ is the ``oldest ancestor'' of the various vertices, or descendants in the tree $\mathcal{T}$.
As a sub-set, the root contains the whole set of elements, i.e. $v_{\Omega}\equiv \Omega$.
For any sub-set $v\in \mathcal{T}$, $\bigtriangleup_v^{\mathcal{T}}$ denotes the set of direct descendants of $v$.
A sub-set $v$ is at the $l$-th level (or depth) of the hierarchy if $l$ is the topological distance from $v$ to $v_{\Omega}$.
When there is no confusion, we simplify the notation to $\bigtriangleup_v$, i.e. by omitting the reference to $\mathcal{T}$.

Consider a  network of nodes $i$ and links  (weighted or not) $w_{ii'}$. 
Here, the terms {\it elements} and {\it nodes} are used interchangeably; both, referring to the entities denoted by $i$.
Traditionally, the community structure of a network is represented by a node partition.
In many cases, these communities present a hierarchical organization.
In particular, if the hierarchy is constituted by sub-communities within communities, then the structure can be mapped to a hierarchical partition $\mathcal{T}$. 
Depending on the context, $\mathcal{T}$ is referred to as a tree, as a hierarchical community structure, or simply as a hierarchy; i.e.~the terms are used interchangeably.
Each sub-set $v\in\mathcal{T}$ corresponds to one and only one sub-community of the network hierarchical community structure (see Fig.~\ref{fig:1}a).
The root $v_{\Omega}$ represents the set of all  nodes in the network.
The children $u\in\Delta_v$ correspond to a partition of the sub-community $v$ into sub-communities $u$.
The leaves of $\mathcal{T}$ are the smallest sub-communities of the network.
Finally, each sub-community $v\in\mathcal{T}$ has an associated sub-network with links $w_{ii'}^{(v)}$, between the pair of nodes $i,i'\in v$.

\subsection{Uncertainty Reduction}

In this section, the definition of the {\em hierarchical mutual information} is motivated.
Only Shannon-based information measures are used throughout the rest of the paper~\cite{cover2006elements}.

Consider how the uncertainty about the identification of a specific node $i$ is reduced when going down a tree 
$\mathcal{T}$.
As the root $v_{\Omega}\in\mathcal{T}$ represents the set of all nodes, to look for a specific node $i$ requires checking $\cong \log_2 |v_{\Omega}|$ binary choices.
In other words, the uncertainty is reduced by $\ln |v_{\Omega}|$ nats when a node $i$ is unequivocally identified~(a nats is a unit of information equals to $1/\ln 2\approx 1.44$ bits), and there is no uncertainty left.
Sometimes the information pointing towards a specific node is not precise, and the uncertainty reduction is not complete.
For example, if node $i$ is specified to be in the sub-community $v$, the uncertainty reduction is $\ln|v_{\Omega}|-\ln|v|=-\ln( |v|/|v_{\Omega}|)$ nats, and  $\ln |v|$ nats of uncertainty still remains.

Traversing a hierarchy along descendants is similar to a sequential reduction of  uncertainty.
More specifically, it is possible to write
\begin{eqnarray}
\label{eq:1}
-\ln 1/|v_{\Omega}| 
&=&
-\ln |v_1|/|v_{\Omega}|
-\ln |v_2|/|v_1|
-
... \nn
&&
... -\ln |v_l|/|v_{l-1}|
    -\ln|v_{l+1}|/|v_l| ... \nn
&&
.. -\ln |v_{L_i}|/|v_{L_i-1}|
-\ln 1/|v_{L_i}|,
\end{eqnarray}
where $L_i$ is the deepest level at which node $i$ can be found.
Each term $-\ln |v_l|/|v_{l-1}|$ can be considered ed a conditional uncertainty reduction.
Specifically, how much the uncertainty is reduced when new information is gained (that $i\in v_l$), given that some other information was already available (that $i\in v_{l-1}$).

It is possible to average over nodes $i$ using an appropriate weighted version of the expression in Eq.~(\ref{eq:1}).
More specifically, the average uncertainty reduction along the tree $\mathcal{T}$ is defined as
\begin{eqnarray}
\label{eq:2}
\avrg{H_{\mathcal{T}}}
&=&
\sum_{v_1\in \bigtriangleup_{v_{\Omega}}}
-\frac{|v_1|}{|v_{\Omega}|}\ln \frac{|v_1|}{|v_{\Omega}|} + ...
\\
&&
...
+\sum_{v_l\in \bigtriangleup_{v_{l-1}}}
-\frac{|v_l|}{|v_{l-1}|} \ln \frac{|v_l|}{|v_{l-1}|} + 
... \nn
&&
+ \sum_{i \in v_{L}} -\frac{1}{|v_{L}|}\ln \frac{1}{|v_{L}|}.
\nonumber
\end{eqnarray}
where, for simplicity, we wrote the equation for the particular case in which all branches of the tree $\mathcal{T}$ have the same length, i.e. $L_i = L$ for all $i\in \Omega$. The general case is introduced later in sub-section~\ref{ss:theo:hmi:def}.
In Eq.~\ref{eq:2}, every reduction step is weighted by the fraction of nodes that are found by following the corresponding branch of the tree $\mathcal{T}$.
Using similar ideas, the hierarchical mutual information is defined in the next section.

\subsection{The Hierarchical Mutual Information}
\label{ss:theo:hmi:def}

In community detection problems, it is customary to quantify the similarity between two inferred community structures using the mutual information between the corresponding node partitions~\cite{danon2005comparing,lancichinetti2009community,fortunato2010community}.
Here, the goal is to introduce the hierarchical mutual information to quantify the similarity between two hierarchical partitions, or trees, associated to corresponding hierarchical community structures.

Consider two trees $\mathcal{T}$ and $\mathcal{T}'$ and two sub-communities $v\in\mathcal{T}$ and $v'\in\mathcal{T}'$, both at the same topological distance, or level $l$, from the roots of their corresponding trees.
It is not necessary for the trees $\mathcal{T}$ and $\mathcal{T'}$, nor the sub-communities $v$ and $v'$ to contain the same elements.
Let $\mathcal{T}_v$ represent the sub-tree of root $v$ obtained from $\mathcal{T}$.
The analogous holds for $\mathcal{T}'_{v'}$.
The hierarchical mutual information between the sub-trees $\mathcal{T}_v$ and $\mathcal{T}'_{v'}$ is denoted by $I(\mathcal{T}_v;\mathcal{T}'_{v'})$.
By definition, it is assumed that 
$I(\mathcal{T}_v;\mathcal{T}'_{v'})=0$
if either $v$ or $v'$ is a leaf of the corresponding tree.
Otherwise, $I(\mathcal{T}_v;\mathcal{T}'_{v'})$ is recursively defined by the formula
\begin{eqnarray}
\label{eq:3}
I(\mathcal{T}_v;\mathcal{T}'_{v'})
&:=&
I(\bigtriangleup_v;\bigtriangleup_{v'}|v\cap v') \nn
&&
+ \sum_{ \substack{ u\in\bigtriangleup_v, u'\in\bigtriangleup_{v'} \\ |v\cap v'|\neq 0 } }
\frac{|u\cap u'|}{|v\cap v'|}
I(\mathcal{T}_u;\mathcal{T}'_{u'}).
\end{eqnarray}
In Eq.~\ref{eq:3}, the first term of the r.h.s.~is called the
{\em one step mutual information}, and is defined as
\begin{eqnarray}
\label{eq:4}
I(\bigtriangleup_v;\bigtriangleup_{v'}|v\cap v')
&:=&
H(\bigtriangleup_v|v\cap v') 
+
H(\bigtriangleup_{v'}|v\cap v') \nn
&&
-
H(\bigtriangleup_v\cap \bigtriangleup_{v'}|v\cap v'),
\end{eqnarray}
where $H(\cdot)$ represents the Shannon entropy.
These terms are computed as
\begin{eqnarray}
\label{eq:5}
H(\bigtriangleup_v|v\cap v') \\
:= & &
\left\{
\begin{array}{ll} \nonumber
\displaystyle \sum_{u\in\bigtriangleup_v} 
\textstyle
-
\frac{|u\cap v'|}{|v\cap v'|}\ln \frac{|u\cap v'|}{|v\cap v'|} 
& \mbox{if $|v\cap v'|\neq 0$} \\
0 & \mathrm{otherwise}
\end{array}
\right.,
\end{eqnarray}
and
\begin{eqnarray}
\label{eq:6}
H(\bigtriangleup_v &\cap& \bigtriangleup_{v'}|v\cap v') \\
&:=&
\left\{
\begin{array}{ll}
\displaystyle
\sum_{ \substack{ u\in\bigtriangleup_{v} \\ u'\in\bigtriangleup_{v'} } }
\textstyle
-
\frac{|u\cap u'|}{|v\cap v'|}\ln \frac{|u\cap u'|}{|v\cap v'|}
& \mbox{if } |v\cap v'|\neq 0 \\
0 & \mbox{otherwise}
\end{array}
\right.. \nonumber
\end{eqnarray}
In all cases, the convention $0\ln 0=0$ is adopted.
Finally, the hierarchical mutual information of two full trees  $\mathcal{T}$ and $\mathcal{T}'$ is denoted and defined by
\begin{equation}
I(\mathcal{T};\mathcal{T}')
:=
I(\mathcal{T}_{v_{\Omega}};\mathcal{T}'_{v'_{\Omega}})
\label{eq:7}
\end{equation}
where $v_{\Omega}$ and $v'_{\Omega}$ are the roots of $\mathcal{T}$ and $\mathcal{T}'$, respectively.

Each term involved in $I(\mathcal{T};\mathcal{T}')$ is non-negative, and thus, the hierarchical mutual information is a non-negative quantity.
Also, $I(\mathcal{T};\mathcal{T}')=I(\mathcal{T}';\mathcal{T})$, i.e.~it is a symmetric function of its arguments.
When the trees $\mathcal{T}$ and $\mathcal{T}'$ are just stars, i.e.~a root plus one generation of descendants, it is possible to think of them as standard partitions.
In this case, the hierarchical mutual information reduces to the standard mutual information. 

Note, the hierarchical mutual information is not a measure of the similarity between the corresponding final partitions of the nodes at the leaves of the trees (except when both trees are stars).
Rather, it is a summation of weighted local one-step contributions, measuring how similar the partitions are at each corresponding point in both trees.
For example, 
if two nodes $i$ and $i'$ are separated at level $l$ in tree $\mathcal{T}$ and at level $l'\neq l$ in tree $\mathcal{T}'$ then, the separation of $i$ and $i'$ contributes with zero to the value of the hierarchical mutual information.

For practical purposes, a {\it normalized} hierarchical mutual information is defined as
\begin{equation}
\label{eq:8}
i(\mathcal{T};\mathcal{T}')=\frac{I(\mathcal{T};\mathcal{T}')}{\sqrt{I(\mathcal{T};\mathcal{T})I(\mathcal{T}';\mathcal{T}')}}.
\end{equation}
We would like to notice the reader that there exists more than one way to normalize the mutual information.
Here, we work with one that takes inspiration from the Cauchy-Schwarz inequality but, future experiments may prove other normalization methods to be more convenient depending on the particular context in which the hierarchical mutual information is used.
The value of $i(\mathcal{T};\mathcal{T}')$ 
lays in the interval $[0,1]$ and attains the maximum 1 if and only if $\mathcal{T}=\mathcal{T'}$, as indicated  by the results of extensive numerical exploration reported in the following sections.
However, 
formal proofs of the previous statements, and the following ones, are still missing.
More specifically, it remains to be proved that: {\it i)} $I(\mathcal{T};\mathcal{T}')\leq I(\mathcal{T};\mathcal{T})$ for all $\mathcal{T}$ and $\mathcal{T}'$ and, {\it ii)} the equality holds if and only if $\mathcal{T}=\mathcal{T'}$.
These statements imply the previous ones, and constitute
desirable properties for a well defined measure of mutual information to have.

\begin{figure}
\includegraphics*[width=.9\linewidth]{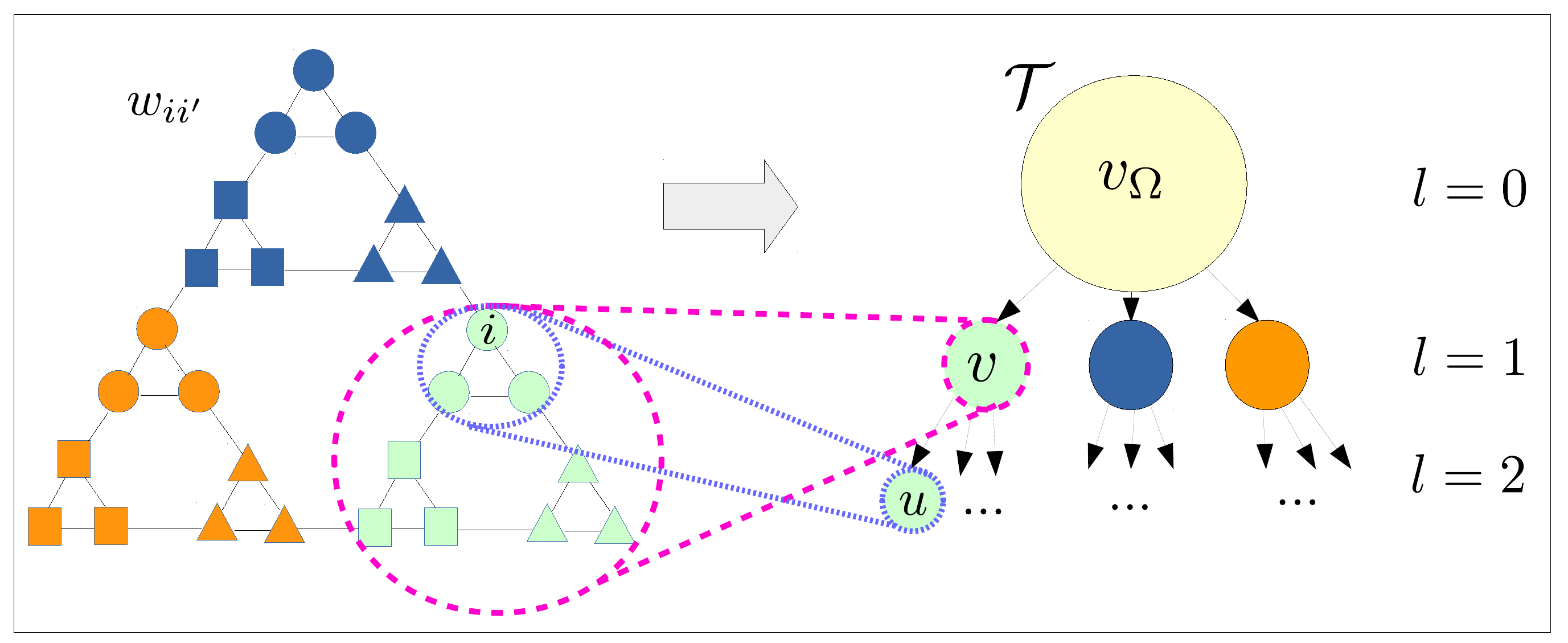}
\put(-215,79){a)}
\\
\includegraphics*[width=.9\linewidth]{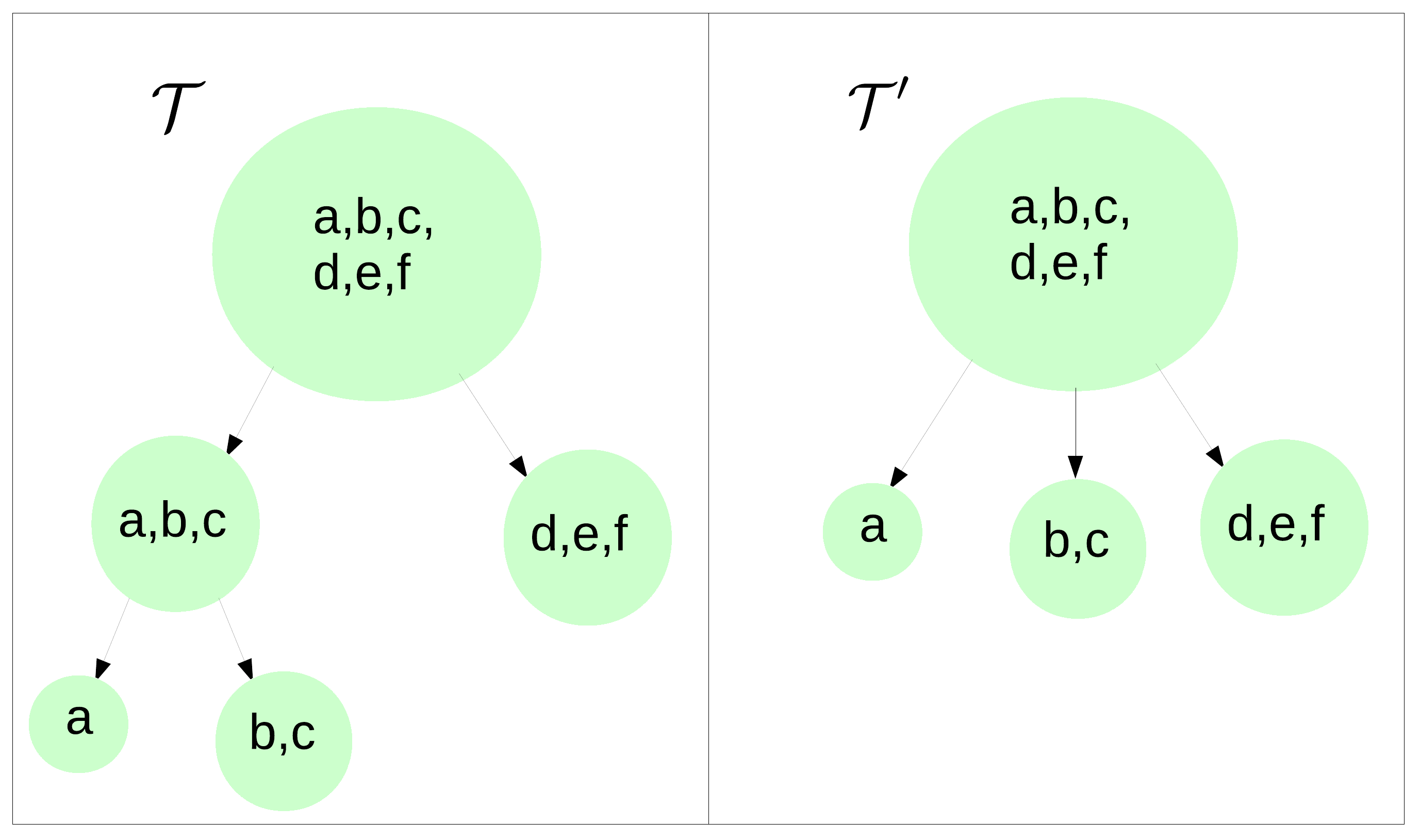}
\put(-215,119){b)}
\put(-105,119){c)}
\caption{
(Color online).
a) Illustration of how a hierarchy of communities obtained from a Sierpinski network $w_{ii'}$ corresponds to a hierarchical partition, or tree $\mathcal{T}$.
The root $v_{\Omega}\in \mathcal{T}$ contains all the nodes of the network $w_{ii'}$, and $v$ represents a sub-community level $l=1$.
In b) and c), two simple hierarchical partitions, or trees, of the same set of nodes, $\{a,b,c,d,e,f,g\}$, are presented.
On the tree $\mathcal{T}$, the node $a$ is separated from the other nodes $\{b,c\}$ at the level $l=2$, while on the tree $\mathcal{T}'$ the separation occurs at level $l=1$.
This difference implies a normalized hierarchical mutual information smaller than one, even if the partition at the bottom of both trees is the same.
}
\label{fig:1}
\end{figure}

To help better understand the hierarchical mutual information, a simple example is worked out explicitly.
Consider the set of nodes $\{a,b,c,d,e,f\}$, and the two hierarchical partitions $\mathcal{T}=\{\{\{a\},\{b,c\}\},\{d,e,f\}\}$ and $\mathcal{T}'=\{\{a\},\{b,c\},\{d,e,f\}\}$ (see Figs.~\ref{fig:1}b~and~\ref{fig:1}c).
Here, $v_{\Omega}=v'_{\Omega}=\{a,b,c,d,e,f\}$.
Also,  $\Delta_{v_{\Omega}}=\{\{a,b,c\},\{d,e,f\}\}$ and $\Delta_{v'_{\Omega}}=\{\{a\},\{b,c\},\{d,e,f\}\}$.
In the tree $\mathcal{T}$, there is an intermediate sub-community $\{a,b,c\}$ which is not on the other tree $\mathcal{T}'$.
As a consequence, the one-step mutual information at level $l=1$ is $I(\bigtriangleup_{v_{\Omega}};\bigtriangleup_{v'_{\Omega}}|\{a,b,c,d,e,f\})\cong 0.693$ (see Eq.~\ref{eq:4}).
All other terms corresponding to levels $l>1$ contribute with zero because they involve leaves.
This is because the tree $\mathcal{T}'$ is just a star which has only one level.
Adding all together,  $I(\mathcal{T};\mathcal{T}') \cong 0.693$.
On the other hand, the {\it self-hierarchical mutual informations} are $I(\mathcal{T};\mathcal{T})\cong 1.011$ and $I(\mathcal{T}';\mathcal{T}') \cong 1.011$.
Therefore, the normalized hierarchical mutual information yields $i(\mathcal{T};\mathcal{T}')\cong 0.685$; a value smaller than one.
In other words, these trees share only a fraction of the information they contain.
This holds in spite that the partitions at the bottom are the same for both trees.

To facilitate future research, collaboration and scientific reproducibility, we provide  Python~\cite{python} code implementing the hierarchical partition data-structure and the hierarchical mutual information function, as an open-source package~\cite{hierpart}.
The example of 
Figs.~\ref{fig:1}b~and~\ref{fig:1}c is provided
in the Python package.

\section{Results}
\label{sec:resu}

\subsection{Testing the Hierarchical Mutual Information in Artificial Hierarchies}
\label{sec:resu:arthier}

Before focusing on the hierarchical community structures 
of networks, we analyze the behavior of the hierarchical mutual information when used to compare artificially generated hierarchical partitions.
More specifically, 
hierarchies composed of binary trees $\mathcal{T}$ containing $N=2^L$ elements $i$, $L$ levels, and $2^{L+1}-1$ sub-communities including the root.
Each tree has one element $i$ per sub-community at the bottom level $l=L$, two elements per sub-community at  the previous level $l=L-1$, and so on until it has $N$ elements at the root.

In the experiments, the original trees are compared against correspondingly randomized ones.
The idea is to show how the normalized hierarchical mutual information decays with respect to the level of randomization.
Two different randomization procedures are used.

\begin{figure}
\includegraphics*[width=.9\linewidth]{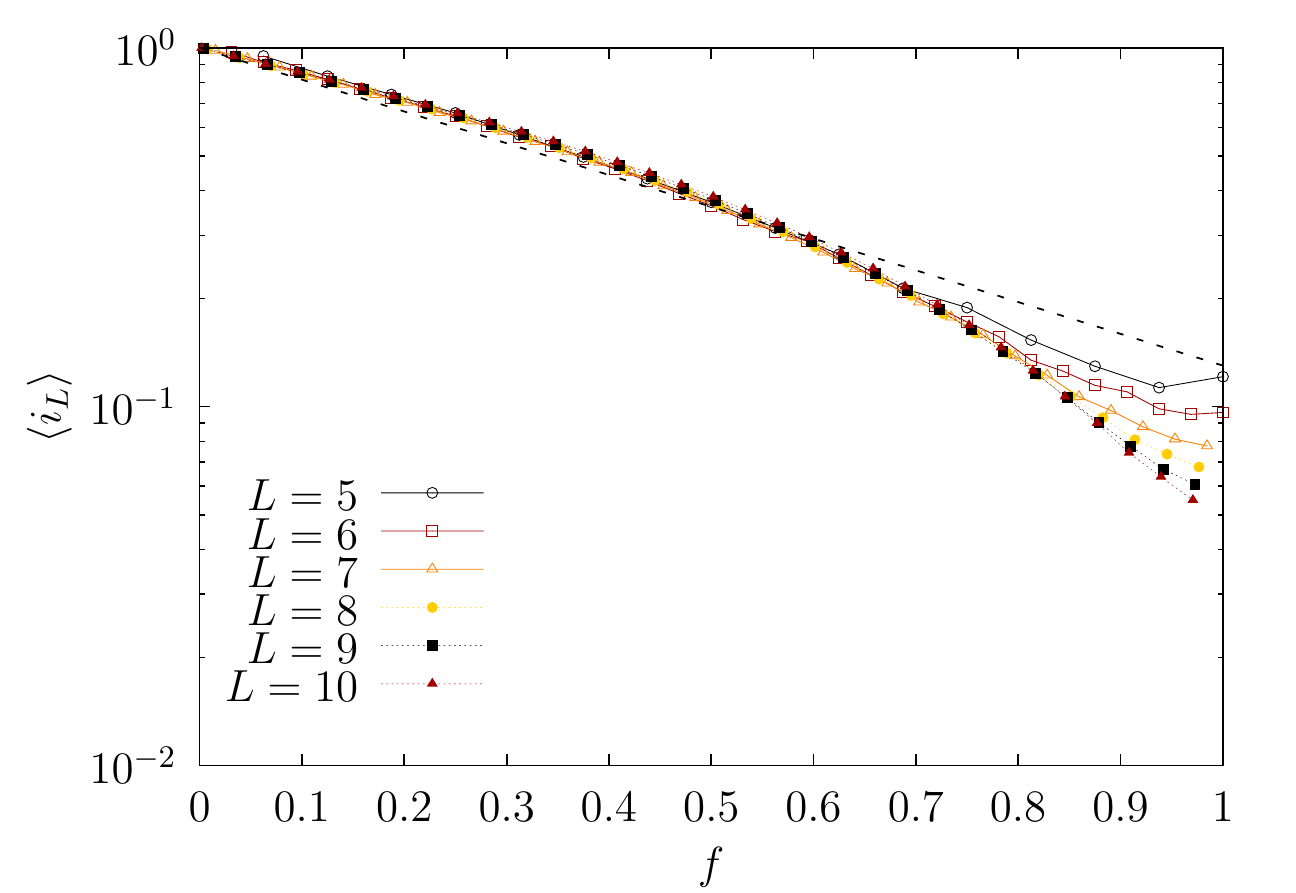}
\caption{
(Color Online).
The normalized hierarchical mutual information, $\avrg{i_L}$, comparing hierarchical partitions represented by binary trees with $L$ levels, and corresponding randomized partitions with a fraction $f$ of the elements shuffled at random.
The average is computed over 100 realizations of the shuffling procedure, and the different curves correspond to trees with different number of levels $L$.
The black dashed line corresponds to an exponential fit, $\avrg{i_L}=\exp(-f/f_0)$ with $f_0=0.490\pm 0.004$ and $R^2=0.968$, for the case $L=10$.
Error bars and standard-deviation bars are not plotted for clarity.
}
\label{fig:2}
\end{figure}

In the first randomization procedure, 
pairs of elements are randomly chosen from the tree, and consecutively swapped until a fraction $f$ of them is affected.
This is called the {\it basic} randomization procedure.
In Fig.~\ref{fig:2}, the average normalized hierarchical mutual information $\avrg{i_L}$
is plotted vs the fraction $f$ of randomized elements.
The average is computed over 100 repetitions of the randomization procedure, for each value of $f$ and $L$.
Notice, $\avrg{i_L}$ decays approximately in an exponential way with respect to $f$; further, it is almost independent of $L$ except for large values of $f$, where finite size effects become important.
In particular, when the hierarchy is fully randomized, i.e.~$f=1$, the $\avrg{i_L}$ is non-zero.
Although {\it a priori} this may  be attributed to an error, it is indeed an expected result for finite size hierarchies: random coincidences produce a non-zero amount of shared information.
A similar result is known to hold for the traditional mutual information~\citep{meilua2007comparing}.

\begin{figure}
\includegraphics*[width=.9\linewidth]{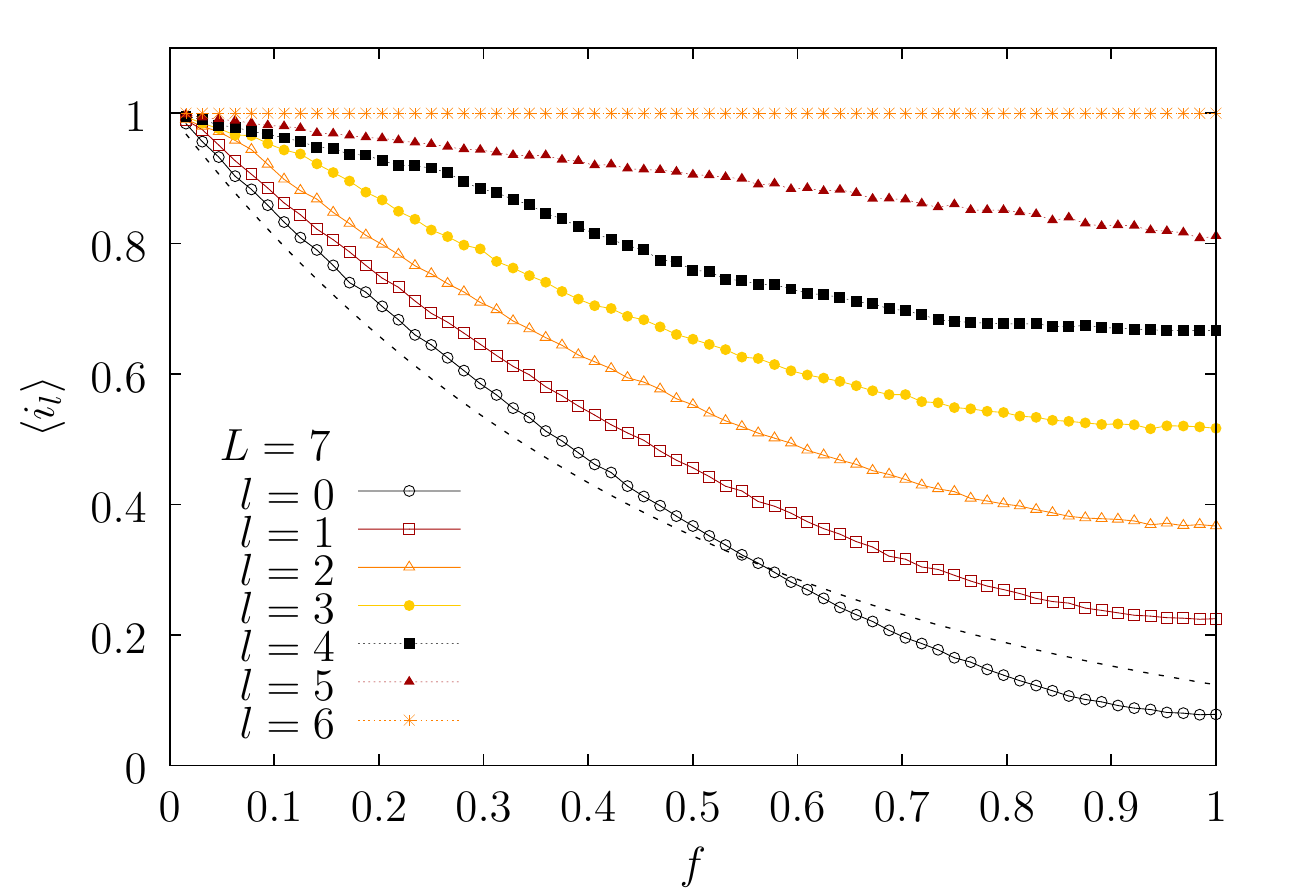}
\caption{
(Color Online).
The normalized hierarchical mutual information comparing hierarchical partitions represented by binary trees with $L=7$ levels, and corresponding randomized partitions where a fraction $f$ of the elements are randomly shuffled.
The randomization procedure preserves the element classification of levels $0,1,...,l-1$, but affects the rest of the levels (different curves with symbols).
The dashed line is the best fit of an exponential decay 
$\avrg{i_l}=\exp(-f/f_0)$, with $f_0=0.478\pm 0.008$ and $R^2=0.978$, for $l=0$.
For clarity, error bars and standard-deviation bars are not shown.
}
\label{fig:3}
\end{figure}

In the second procedure, the elements are also shuffled by swapping pairs chosen at random.
However, a given pair is swapped only if both elements belong to the same sub-community at depth $l$.
In other words, the randomization procedure preserves the classification of the elements at the levels $0,1,...,l-1$, while in the subsequent levels $l,l+1,...,L$, the original classification is destroyed.
Again, the swapping procedure runs until a fraction $f$ of the elements is affected.
This second procedure is called the {\it level-preserving} randomization procedure.
In Fig.~\ref{fig:3}, the average normalized hierarchical mutual information $\avrg{i_l}$ is plotted as a function of $f$ for the level-preserving randomization procedure.
Here, experiments are repeated for different values of $l$ and fixed $L=7$.
Averages are computed as it was done with the basic randomization procedure.
In line with the previous result of Fig.~\ref{fig:2}, $\avrg{i_l}$ also decreases with $f$ following approximately an exponential decay.
Now the greater is the shuffling level $l$, the slower is the decay.
In particular, for $l=6$ no decay at all is observed, i.e.~$\avrg{i_l}=1$ for all $f$.
This is expected because trees have $L=7$ levels and only one element per sub-community at the bottom level, which do not contribute to the hierarchical mutual information.

\subsection{Comparing Community Detection Algorithms on Artificial Networks}
\label{sec:resu:artnets}

\subsubsection{Community Detection Methods}
\label{sss:cdm}

One of the  interesting applications of the normalized hierarchical mutual information is  comparing the results yielded by different community detection methods.
In this paper, three community detection methods are compared: {\it Infomap}~\cite{rosvall2011multilevel}, which find a hierarchy of communities through the minimization of the description length of the path traversed by a random walker; the {\it Hierarchical Stochastic Block Model} method (HSBM)~\cite{peixoto2014hierarchical}, which fits a hierarchy of stochastic block models to the network topology; a {\it Recursive Louvain} method (RL), which  recursively splits the network into a hierarchy of network modules using, at each step, the well-known Louvain community detection algorithm~\cite{blondel2008fast}.
In what follows, the relevant aspects of the different methods are considered in more detail.

Infomap return hierarchies that are consistent with a \textit{divisive} algorithm, i.e.~the branches of the corresponding trees may have different depths.
The algorithm itself uses both approaches, repeatedly.
Communities are split and merged until a minimum description length is attained.
In the hierarchies obtained by this method, the leaves have one and only one node.
For the sake of comparison with the other methods, these communities of size equal to one are ignored, except if same level communities of size larger than one exists.

At difference with Infomap,
the HSBM merge nodes to generate super-nodes or communities, which are further merged to obtain the communities at the contiguous higher level, and so on.
As a consequence, all the branches of the returned trees have the same depth.
Moreover, the HSBM may return trees containing sub-communities with descendants but no further sub-divisions, i.e., sub-communities with only one child.
Although the hierarchies produced by the HSBM can be compared using the hierarchical mutual information -- as they are hierarchical partitions -- the comparison are not fully appropriate.
This is because the hierarchical mutual information is based on a divisive approach, while the HSBM is based on an agglomerative approach.
The experiments involving the HSBM show how important is the difference between both kind of approaches.

The recursive application of the Louvain method is a mixed agglomerative-divisive algorithm.
The standard Louvain method is an agglomerative algorithm; a community structure is obtained by merging modules until
the {\it modularity}~\cite{newman2006modularity}
of the partition, denoted by $Q$, reaches a maximum value~\cite{blondel2008fast}.
On the other hand,
the recursive use of Louvain presented here, is a divisive method.
More specifically, given a network $w_{ii'}$ (defining the level $l=0$), a standard Louvain method is applied to obtain a partition into sub-communities $v$ at level $l=1$.
Then, Louvain is applied again on each sub-community, to split each sub-network $w_{ii'}^{(v)}$ into sub-communities at level $l=2$, and so on.
In this way, a tree $\mathcal{T}$ is generated.
The division of a particular sub-community stops when the standard Louvain returns a modularity $Q\leq 0$.
Importantly, the Louvain method is not deterministic, leading to stochastic differences from run to run.
Two important points have to be stressed: 
First, the use of Louvain is circumstantial, any other modularity maximization procedure would produce similar results.
Second, the idea of a recursive application of a modularity based community detection algorithm is not new, and more elaborate algorithms do exist~\cite{pons2011post,zhang2014scalable,macmahon2015community}.
However, here RL is chosen for its simplicity.
Our main goals are: to show how the hierarchical mutual information behaves, and to illustrate how it can be used, without aiming to find the best community detection method.

\subsubsection{Artificial Hierarchical Networks}\label{sss:ahn}

In order to analyze the performance of the different community detection methods, in this section they are run on specific networks.
Here, two well-known benchmark network models are used to generate the networks necessary for the experiments.
In principle, these network models are able to generate network samples with underlying hierarchical community structures.
Clearly, the specific characteristics of the generated networks depend on the parameter values chosen.

\begin{figure}
\includegraphics*[width=1\linewidth]{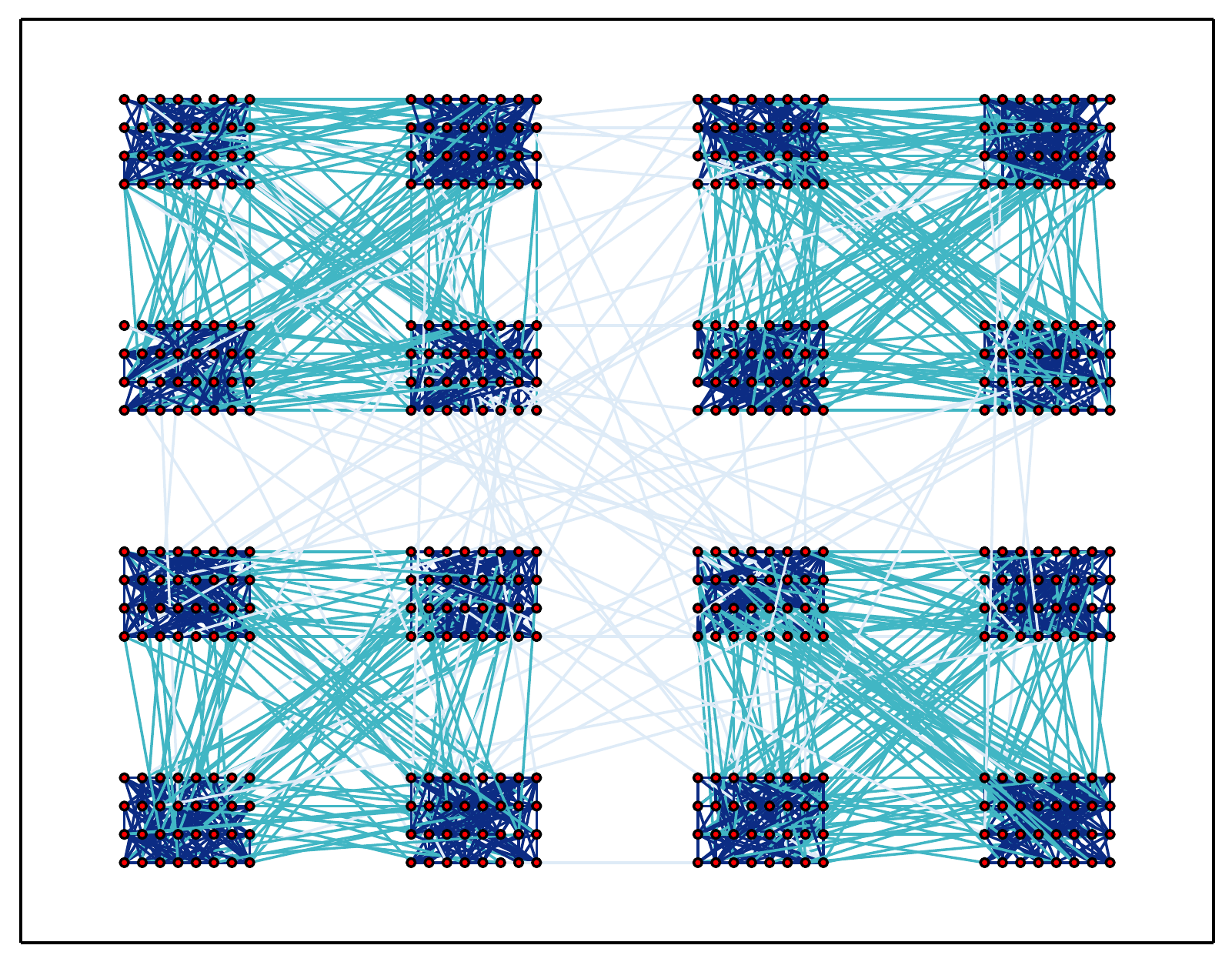}
\caption{
(Color Online).
A sample network obtained from the hierarchical planted network model (HPM), for $K_0=0.25,K_1=2$ and $K_2=8$.
The network contains $N=512$ nodes, 4 big communities of 128 nodes each, and 16 small communities of 32 nodes.
The darkest links connect pairs of nodes sharing the same small community at level $l=2$, the  links of intermediate brightness connect nodes sharing intermediate communities at level $l=1$, but not sharing the same small communities.
Finally, the brightest links connect pairs of nodes at level $l=0$, but not nodes sharing communities at levels $l=1$ and $l=2$.
}
\label{fig:4}
\end{figure}

The first network model is the hierarchical planted partition model (HPM)~\cite{lancichinetti2009detecting}, a generalization of the planted partition model~\cite{condon2001algorithms} where the network obtained is hierarchically arranged.
In this model, $N$ nodes are connected according to a hierarchical structure of $L$ levels and a branching factor $B$.
For practical purposes, we chose $N=512$ nodes, $L=3$ levels and a branching factor $B=4$~(see Fig.~\ref{fig:4}).
At the root level, $l=0$, all nodes belong to the same community.
At level $l=1$, there exist $B=4$ communities with $128$ nodes each.
Consecutively, at the final level $l=2$, there are $B^2=16$ communities, with 32 nodes each.
Each node has an average of $K_l$ links to nodes exclusively within the community they belong at level $l$, i.e.~$K_2$, $K_1$, $K_0$ to other nodes in the same communities at levels 2, 1, 0, respectively. 
Therefore, the total average degree of the nodes is $\avrg{k}=K_0+K_1+K_2$.
In principle, networks sampled from the HPM have the expected hierarchical community structure whenever $K_0<K_1<K_2$~\cite{lancichinetti2009detecting}.

The second network model consists of \textit{Sierpinski} networks with $L$ levels.
Fig.~\ref{fig:1}a illustrates a Sierpinski network with $L=3$.
These networks have a natural self-similar and hierarchical modular structure.
A Sierpinski network with a single level is just a clique with 3 nodes, i.e. a triangle.
A network of this type with $L+1$ levels is obtained by by replacing each node of a Sierpinski network with $L$ levels by a clique of size 3.
It is worth to point out that a Sierpinski network with $L$ levels has $N(L)=3^L \sim e^L$ nodes, $M(L)=3[M(L-1)+1] \sim e^L$ links, and its average degree $\avrg{k} \to 3$ when $L\to \infty$.
To make the analysis more interesting, a fraction $f$ of the links in the Sierpinski networks are randomly rewired.
The rewiring procedure is well-known~\cite{maslov2002specificity}.
Essentially, successive pairs of links, each of which is chosen at random, swap the nodes at their extremes until a fraction $f$ of the links is affected.
In this way, there is a well-defined hierarchy of communities for $f=0$, which is progressively blurred out as $f$ increases.

In the following sections, the different community detection methods, and these two network models are combined into a set of experiments analyzed using the normalized hierarchical mutual information.

\subsubsection{Hierarchical-fidelity}
\label{ss:art_fid}

Each network model has an associated natural or \textit{reference} hierarchy, denoted as $\mathcal{T}^*$.
Specifically, the reference hierarchy for the Sierpinski and HPM models are shown in Figs.~\ref{fig:1}a and \ref{fig:4}, respectively.
The hierarchies identified by the community detection methods are not necessarily equal to the reference ones, and in some cases, they don't even resemble it.
The degree of fidelity of the community detection method measures how similar are the identified communities to the reference ones.
Formally, given a community detection method, a network sample $w_{ii'}$ and a reference hierarchy $\mathcal{T}^*$, the {\it hierarchical-fidelity} -- or simply, {\it fidelity} -- of the community method is defined as the average normalized hierarchical mutual information,
$\avrg{i(\mathcal{T}^*;\mathcal{T})}$.
The average is computed summing over an ensemble of hierarchies $\left\{ \mathcal{T} \right\}$, obtained by repeatedly identifying the hierarchical community structure of the network $w_{ii'}$, using the chosen community detection method.
In the results shown, the ensemble $\left\{ \mathcal{T} \right\}$ was composed by 100 hierarchies.
Furthermore, the fidelity is averaged by sampling 100 networks from each network model.
The procedure is repeated for different values of the network models parameters, and using the different community detection methods.

For the case of the HPM, two different model re-parameterizations are used.
In one case the whole network structure change simultaneously, while in the other case, only one level is affected~\cite{lancichinetti2009detecting}.
More specifically, in case 1 all parameters $K_0 = 7.75 \mu + 2$, and $K_1 = 6 \mu +2$ are linearly re-parameterized by $\mu\in [0,1]$, while $K_2$ is kept constant. 
In case 2, the parameters $K_0 = K_2$ = 8 are kept constant, while $K_1 = 8 \mu +4$ changes linearly with $\mu$.
For the case of the Sierpinski network model, the parameter is the fraction of randomized links, $f$, as  mentioned in Section \ref{sss:ahn}.
In what follows the results are presented and commented.

\begin{figure*}
\includegraphics*[width=.32\linewidth]{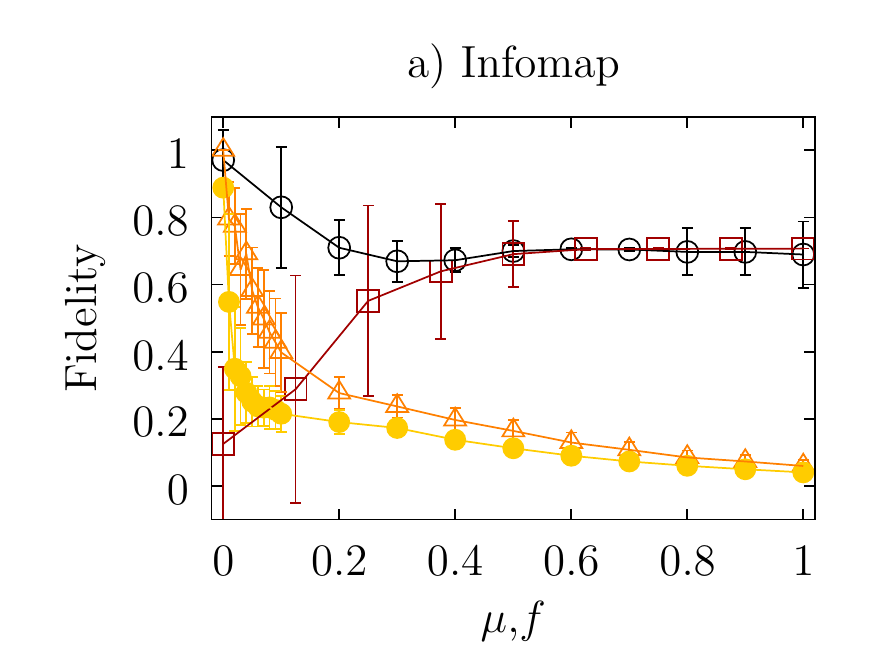}
\includegraphics*[width=.32\linewidth]{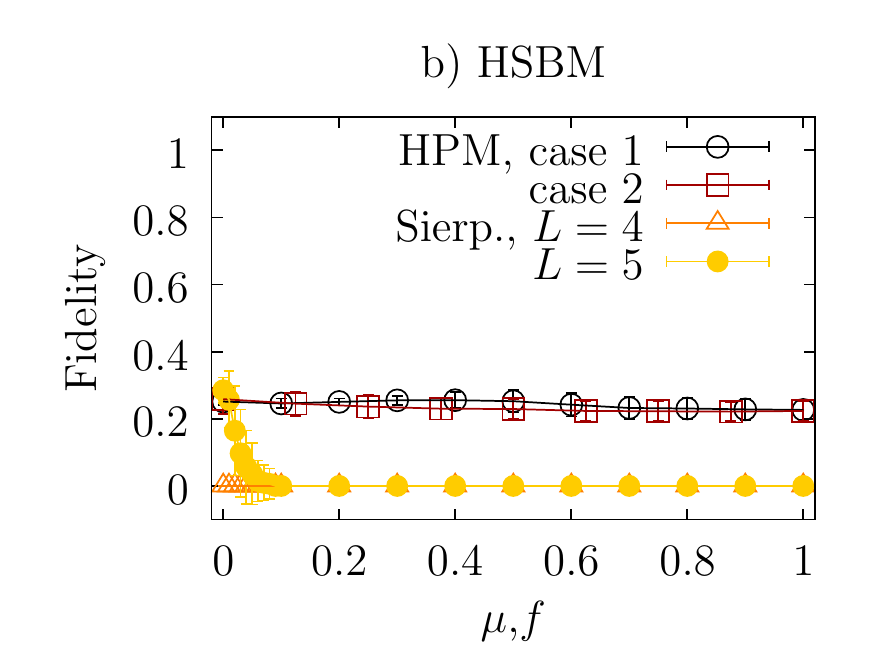}
 \includegraphics*[width=.32\linewidth]{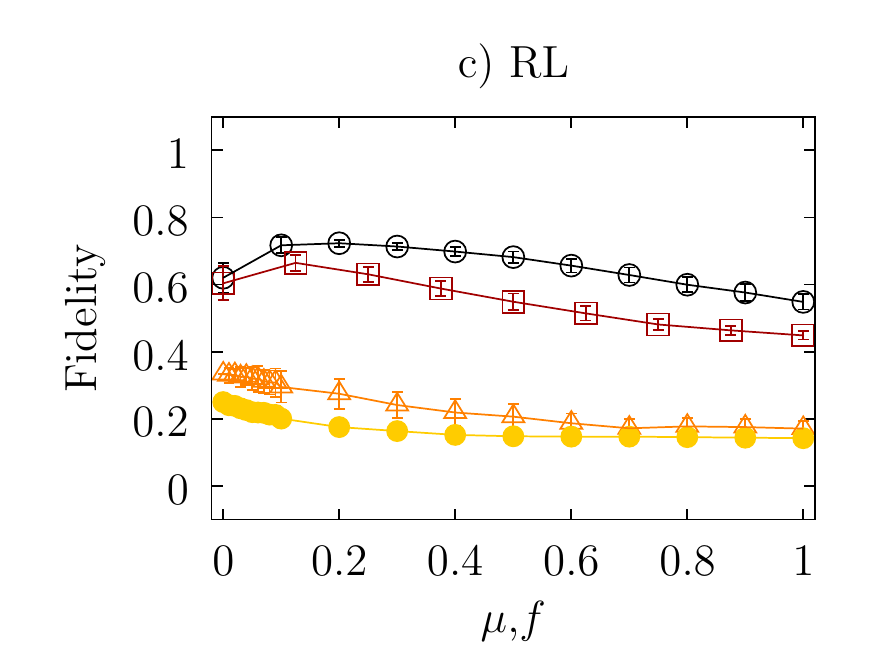}
\caption{
(Color Online).
The fidelity compares the hierarchical community structure of the networks generated with the models against corresponding reference hierarchies.
The topology of the generated networks changes as a function of the parameter $\mu\in [0,1]$.
For the HPM, case 1, $\mu$ parameterizes the network model according to $K_0(\mu)=7.75\mu+0.25$, $K_1(\mu)=6\mu+2$ and $K_2=8$.
Similarly, in case 2, $K_0=K_2=8$ and $K_1(\mu)=8\mu+4$.
For the Sierpinski model, $f\in [0,1]$ is the fraction of rewired links and $L$ is the number of network levels.
Each panel corresponds to one of the community detection methods discussed in the text: a) Infomap, b) HSBM, and c) RL.
In all cases, the bars represent standard-deviations around the mean.
}
\label{fig:5}
\end{figure*}

First, the results of the fidelity for Infomap are shown in Fig.~\ref{fig:5}a.
In the HPM, case 1,
Infomap detects the reference hierarchy almost always for $\mu=0$, and the fidelity is $\approx 1$.
On the other extreme, at $\mu=1$,
Infomap typically finds a one-level hierarchy composed of 4 communities with 128 nodes.
The 4 communities are the right ones at the level $l=1$, and the fidelity decays to $\approx 1/\sqrt{2} \cong 0.707$.
The decay in the fidelity is expected because all $K_l$ converge to the same value $K_l=8$ when $\mu\to 1$, making the generated network hierarchies less defined.
In case 2, the same scenario occurs for $\mu \ge 1/2$, i.e. the same 4 communities are identified.
On the other hand, for small $\mu$, the structure of the network is  dominated by  links at levels 0 and 2.  
As a consequence, and depending on the particular network realization, Infomap finds a one-level hierarchy with either 1 or $\approx 16$ communities,  resulting in a small fidelity value.
For the Sierpinski networks, the behavior can be more easily interpreted.
For small $f$, Infomap finds  an approximately accurate representation of the exact hierarchy of communities.
However, as $f$ grows, the hierarchy is quickly blurred out and the fidelity decays accordingly.

The findings of the fidelity for the HSBM method are shown in Fig.~\ref{fig:5}b.
For the HPM, the fidelity is almost a constant function of $\mu$, for both cases 1 and 2.
A closer inspection reveals that, typically, the HSBM method splits the network samples into two communities at level $l=1$,  which are then further subdivided, giving rise to a hierarchy with 3 levels.
Interestingly, the identified hierarchies are similar regardless of the value of $K_1$.
Therefore, the	 resulting fidelity is relatively small because the identified hierarchies are significantly different for the reference one.
In essence, the two communities identified at level $l=1$ mean a significant difference with respect to the expected value of 4.
For the case of the Sierpinski model, the HSBM typically detects only one community, except for vanishing $f$ and $L=5$ where the network splits into two big ones.
For this second model, the fidelity is also generally small.
In our view, this occurs because of two characteristics of the HSBM.
On the one hand, the HSBM follows a conservative approach; no divisions are introduced until there is enough statistical evidence to justify them in terms of a hierarchy of stochastic block models.
On the other hand, the HSBM follows a bottom-up approach -- the elements in $\Omega$ are iteratively merged into modules, super-modules and so on, generating a tree $\mathcal{T}$ with all branches of the same topological length -- while the hierarchical mutual information is more appropriate to compare top-down hierarchies (see \ref{sec:intro}).
In this sense, the comparison of the other methods with the HSBM by means of the hierarchical mutual information, highlights the crucial difference between top-down and bottom-up approaches.

Thirdly, the fidelity is computed for the recursive Louvain method and the corresponding results are shown in Fig.~\ref{fig:5}c.
For the HPM, the fidelity is not a monotonic function of $\mu$, instead  it displays a maximum at an intermediate value of $\mu$.
	In general, this method tends to find the right communities at the first level $l=1$.
However, the random fluctuations of the network samples become meaningful information for RL, and therefore it tends to split the networks into more communities than the originally found in the reference hierarchy.
As a consequence, the normalized hierarchical mutual information yields values smaller than one.
However, because the information shared at level $l=1$ is non-trivial and fairly accurate, the normalized hierarchical mutual information is far from being negligible.
On the other hand, for the case of the Sierpinski network model, RL has a poor performance.
In essence, this method finds significantly more communities than expected, even at level $l=1$, resulting in small fidelity values for all $f$.

\subsubsection{Hierarchical-consistency}

In the previous section, it was shown that each community detection method returns hierarchies different from the expected ones; therefore, some  questions arise.
How mutually consistent are the returned hierarchies?
Do these hierarchies represent noise, or represent a specific detected bias?
The following set of experiments addresses these questions.
More specifically, the idea is to analyze how mutually similar, or consistent are the communities detected by the methods.
Formally, the {\it hierarchical-consistency} -- or just, {\it consistency} -- of a  method is defined as the average normalized hierarchical mutual information $\avrg{i(\mathcal{T};\mathcal{T}')}$, where the average is computed over an ensemble of pairs of hierarchies, $\{(\mathcal{T},\mathcal{T}')\}$.
The hierarchies in the pairs are randomly chosen, without repetition, from the ensembles of hierarchies generated in the previous experiments about the fidelity.
The procedure is repeated for each network sample in order to average the consistency.
The whole procedure is repeated for the different network models and corresponding parameters.

\begin{figure*}
\includegraphics*[width=.32\linewidth]{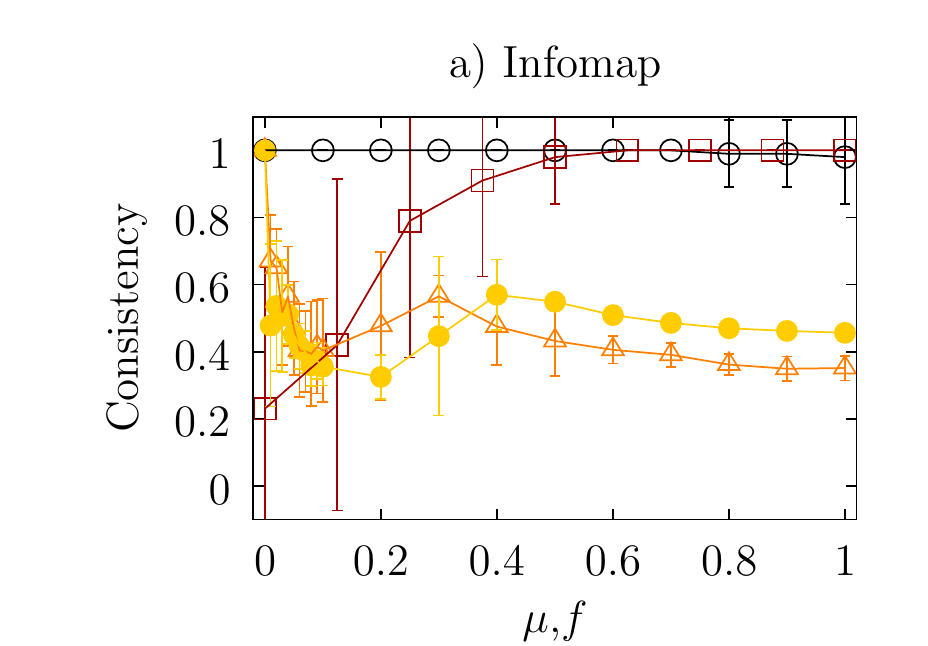}
\includegraphics*[width=.32\linewidth]{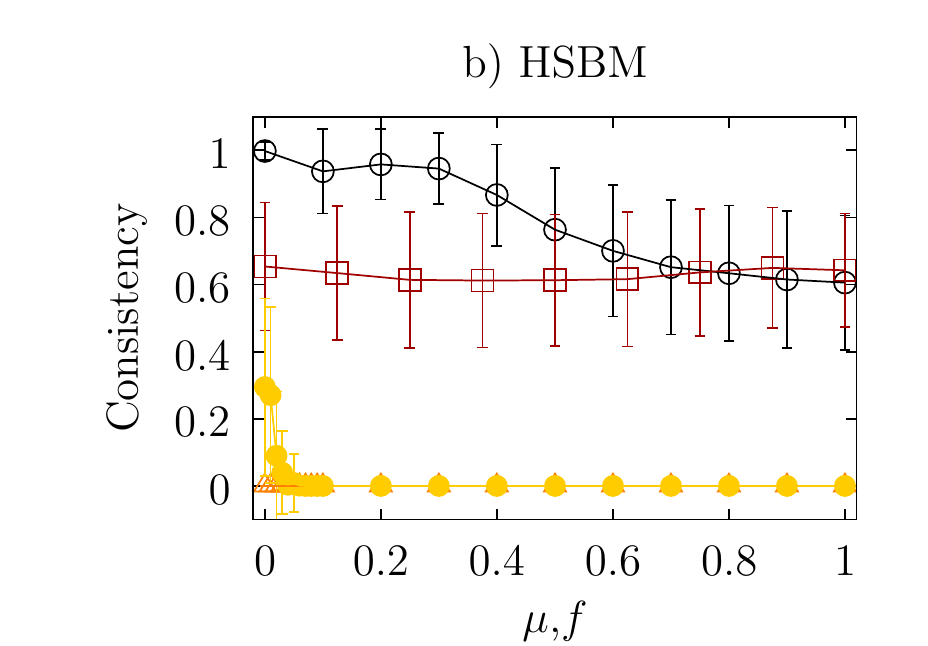}
\includegraphics*[width=.32\linewidth]{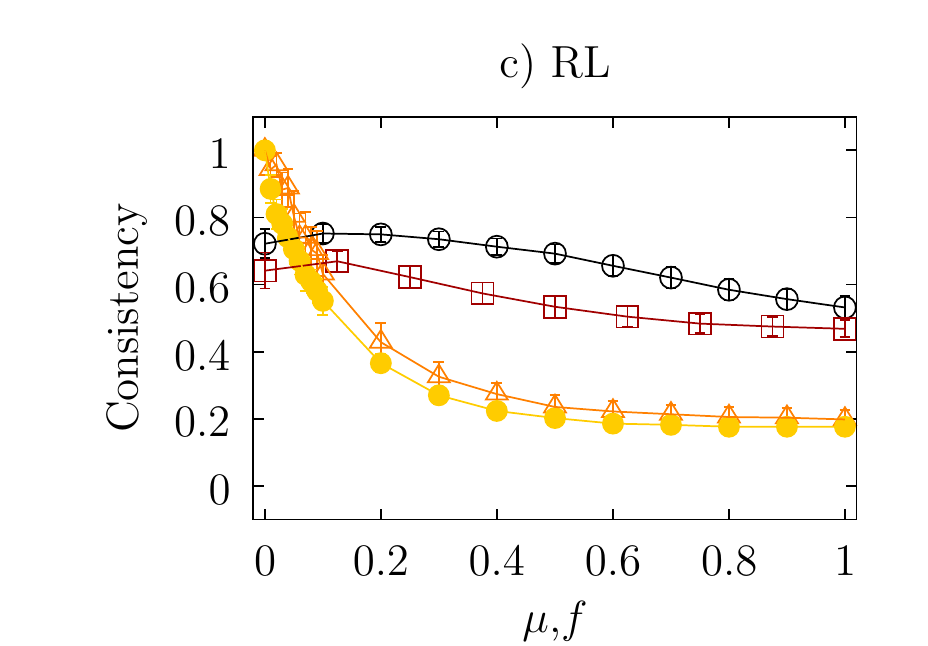}
\caption{
The consistency measures how similar are the different hierarchies obtained from a given community detection method with respect to each other in a specific network sample.
The computations are repeated for several network samples, for each network model, and different values of the parameters $\mu$ and $f$.
See Fig.~\ref{fig:5} for specific details of the simulation parameters.
The panels, (a), (b) and (c) correspond, respectively, to the different community detection methods: Infomap, HSBM and RL.
In all cases, the bars represent standard-deviations around the mean. 
}
\label{fig:6}
\end{figure*}

In Fig.~\ref{fig:6}a, the consistency is analyzed when the hierarchical communities are detected by using Infomap.
For the HPM, case 1 (see Section \ref{sss:ahn}), the consistency is $\approx 1$ for all values of $\mu$.
In other words, in this initial setting, Infomap provides very consistent results always.
For case 2, the fidelity is also close to 1 when $\mu$ is large; however, the consistency becomes small for small $\mu$.
This is expected, as it was already mentioned Infomap's detection is largely bimodal: either it finds one or $\approx 16$ communities depending on the network sample, and these two cases are very inconsistent with each other.
For the Sierpinski networks, the consistency is large when 
$f\approx 0$ and decays to a non-zero value for larger values of $f$.
In other words, network randomization becomes important for large $f$, but still, part of the information captured by Infomap is already contained even in this case. 

The results of the consistency for the HSBM are shown in Fig.~\ref{fig:6}b.
For the HPM, the observed consistency is large in both cases, 1 and 2, despite the small fidelity with respect to the natural hierarchies shown in Fig.~\ref{fig:5}b.
This means that the HSBM return hierarchies similar to each other, but significantly different from the reference one.
More specifically, the returned hierarchies share similarities at level $l=1$, but at the following levels the differences become important -- except for case 1 at $\mu=0$ where the consistency remains $\approx 1$. 
For the Sierpinski network model, the consistency is negligible in most of the range of $f$. 
This is expected because a flat hierarchy conveys no information, and the HSBM typically returns trivial hierarchies for the Sierpinski networks, i.e.~hierarchies with only one community, the root one.
Only for small values of $f$, for the case $L=5$, the consistency is non-zero, but still with small values.
Here, only two communities are identified, agreeing only over a small fraction of the nodes.

The consistency for the RL method is shown in Fig.~\ref{fig:6}c.
For the HPM, the curves look similar to the ones corresponding to the fidelities in Fig.~\ref{fig:5}c.
In essence, the computed hierarchies are very similar to each other, and to the reference hierarchy.
For the Sierpinski network model,
the consistency can be large, even if the fidelity is small.
This means that the detected structure is invariably the same, although different from the reference one.

\subsubsection{Hierarchical-similarity}

It is clear that the different community detection methods return different results.
However, it remains to analyze how similar are the results of one detection method with respect to one another.
To address this point, the {\it hierarchical-similarity} between two community detection methods is defined as the average normalized hierarchical mutual information,
$\avrg{i(\mathcal{T}_1;\mathcal{T}_2)}$.
In shorthand, we speak about the {\it similarity}, and he average is computed over pairs of trees, where the trees $\mathcal{T}_1$ are computed with one of the methods, while the trees $\mathcal{T}_2$ with the other method.
Both set of trees are computed from the same network sample.
Later, the similarity is averaged by sampling networks from the different network models.
The procedure is repeated for each set of chosen values of the corresponding model parameters.
In practice, the network samples and corresponding sampled trees used to compute the fidelities are the ones used to compute the similarities (see Figs.~\ref{fig:5} and \ref{fig:6}).

\begin{figure*}
\includegraphics*[width=.32\linewidth]{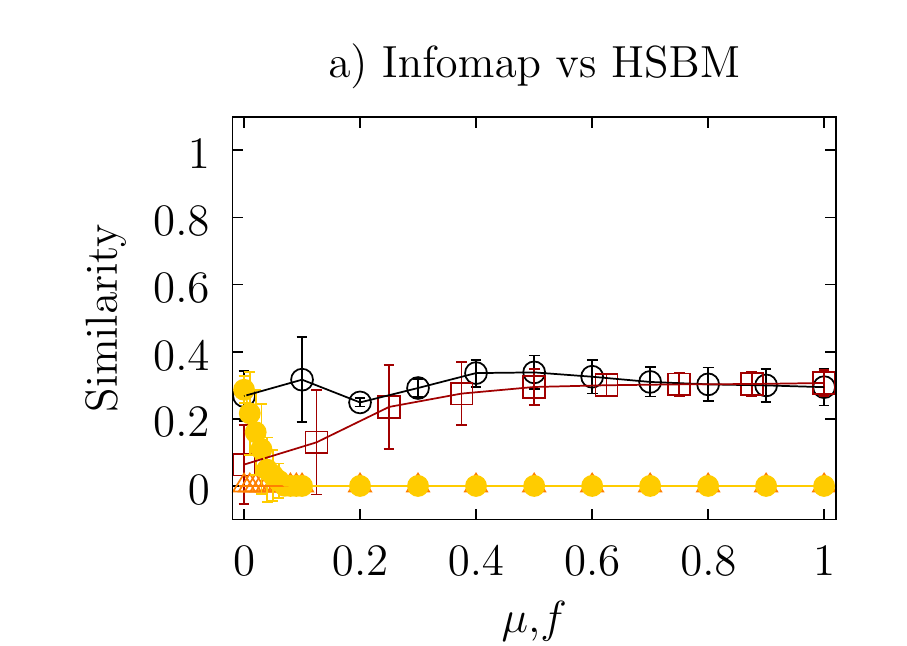}
\includegraphics*[width=.32\linewidth]{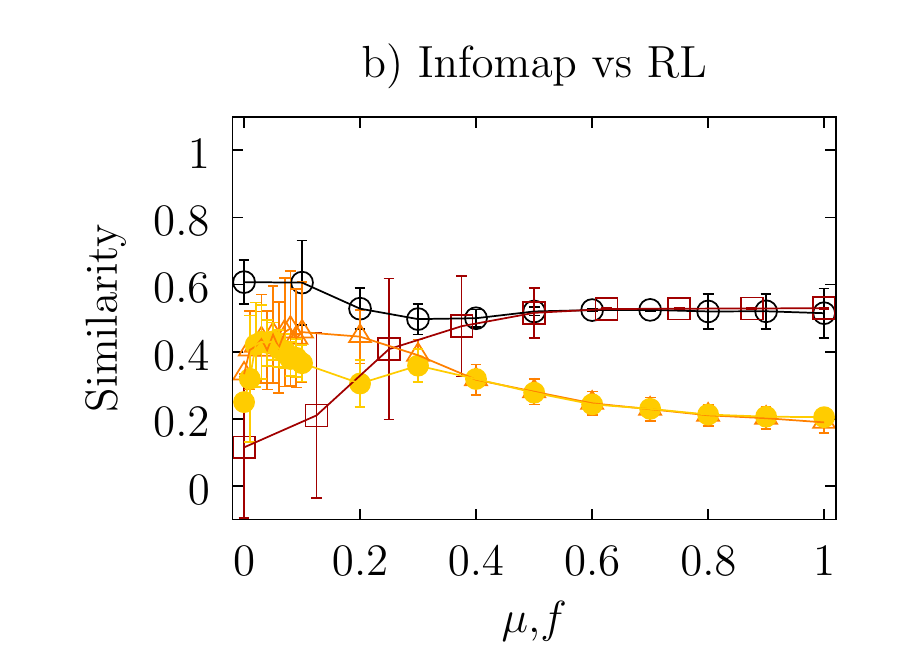}
 \includegraphics*[width=.32\linewidth]{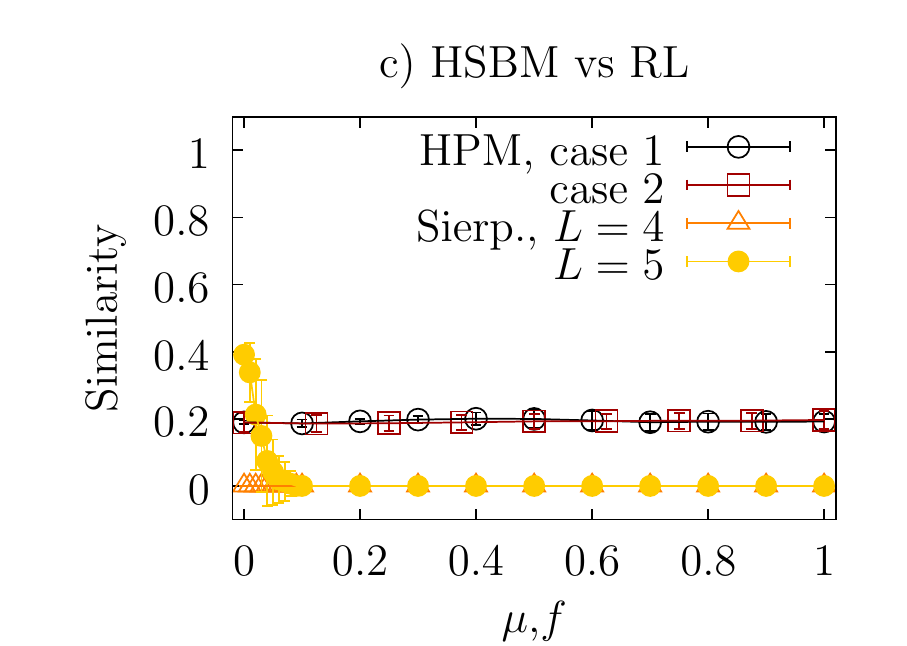}
\caption{
(Color Online).
The similarity compares how similar are the hierarchies obtained by two different community detection methods methods.
Here, we compare: a) Infomap vs HSBM, b) Infomap vs RL, and c) HSBM vs RL.
The hierarchical community structures used to compute the similarities are the same as those used in Fig.~\ref{fig:5}.
Results are shown as a function of the parameters $\mu$ and $f$ of the corresponding network models.
In all cases, the bars represent standard-deviations around the mean.
}
\label{fig:7}
\end{figure*}

Combining the methods of Infomap, HSBM and RL, three different comparisons are possible: Infomap vs.~HSBM, Infomap vs.~RL, and HSBM vs.~RL.
These are presented in Figs.~\ref{fig:7}a,~\ref{fig:7}b and ~\ref{fig:7}c, respectively.
The HSBM method shares a small similarity with the other two.
This is expected, because the other methods lead to relatively large fidelities, while the HSBM does not.

The similarity between Infomap and the RL method is the largest among the three possibilities.
However, the similarity cannot be as large as the consistency.
This is not surprising as Infomap is able to return consistencies as large as 1, while RL is not.
The largest similarity value is $\approx 0.6$, occurring at $\mu=0$ for the case 1 in the HPM.
Also, the similarity is $\approx 0.5$ at $\mu=1$ for both cases, 1 and 2.
For the Sierpinski network, the similarity reaches a maximum value $\approx 0.5$ for small $f$, and it decays slowly up to $\approx 0.2$ for large $f$.

\subsection{Analysis of the Hierarchical Modular Structure of Complex Networks}
\label{sec:resu:empnets}

The experiments of the previous section can be repeated using empirical networks -- as opposed to network models -- except for the computation of fidelity because, a priori, it is not clear which one is the concomitant reference hierarchy.
Notice however, this last possibility is not necessarily impossible for all empirical networks. 
Many empirical networks have associated a hierarchical decomposition that can be used as a ``ground-truth'' about its hierarchical structure.
Let us remark here that by ground-truth we refer to the practical use of the term~\cite{hric2014community}.
For example, the {\it NAICS}~\cite{naics_url} codes for the case of financial networks~\cite{mantegna1999hierarchical,bonanno2003correlation,bonanno2004networks,macmahon2015community}, 
and the {\it Harmonized System}~\cite{baci_wcoomd} for the case of the international trade network~\cite{hidalgo2007product,barigozzi2010multinetwork,caldarelli2012network}.
However, these studies are left open for future research and, in what follows, only consistencies and similarities are analyzed in different empirical networks.

\begin{figure*}
\includegraphics*[width=.32\linewidth]{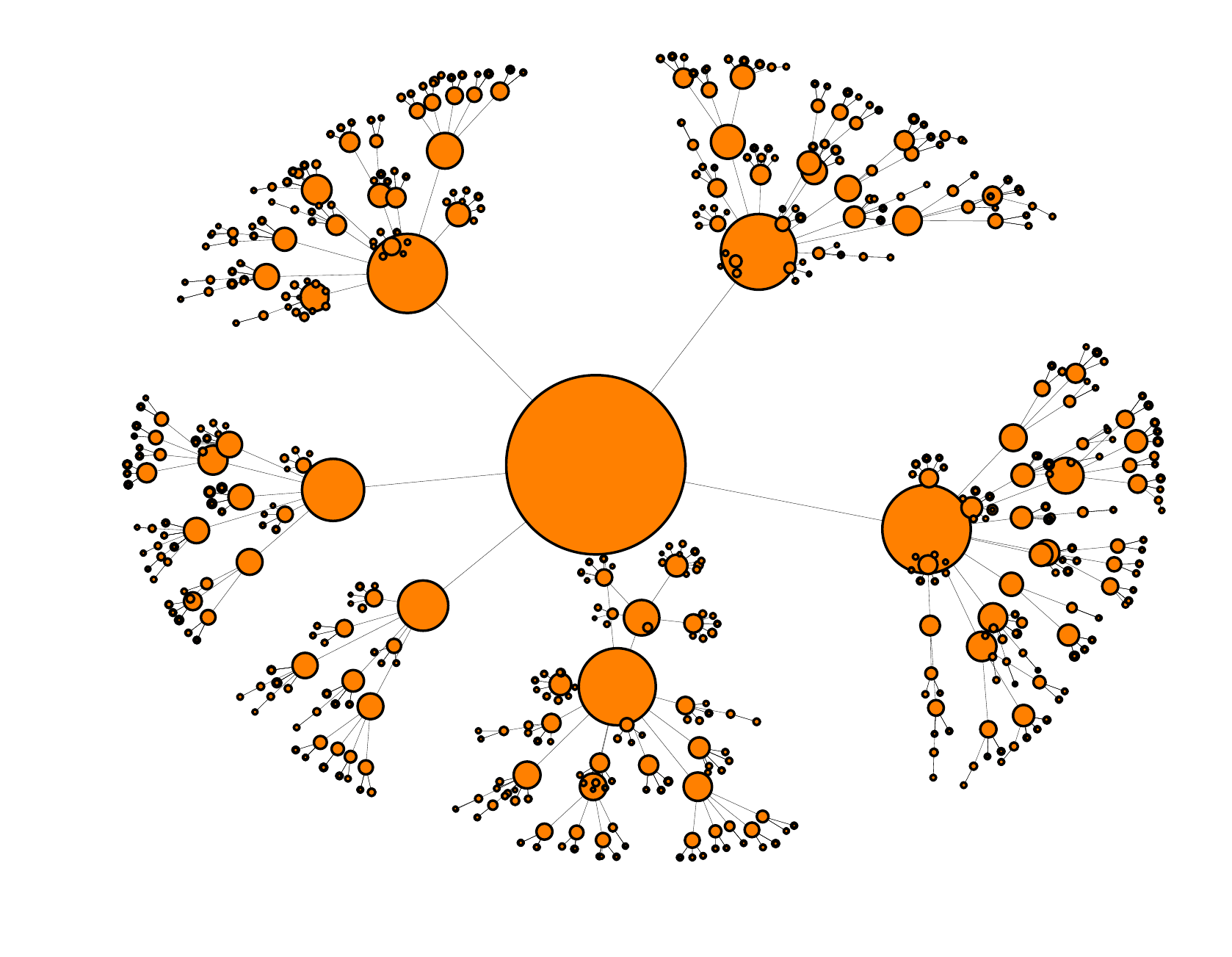}
\includegraphics*[width=.32\linewidth]{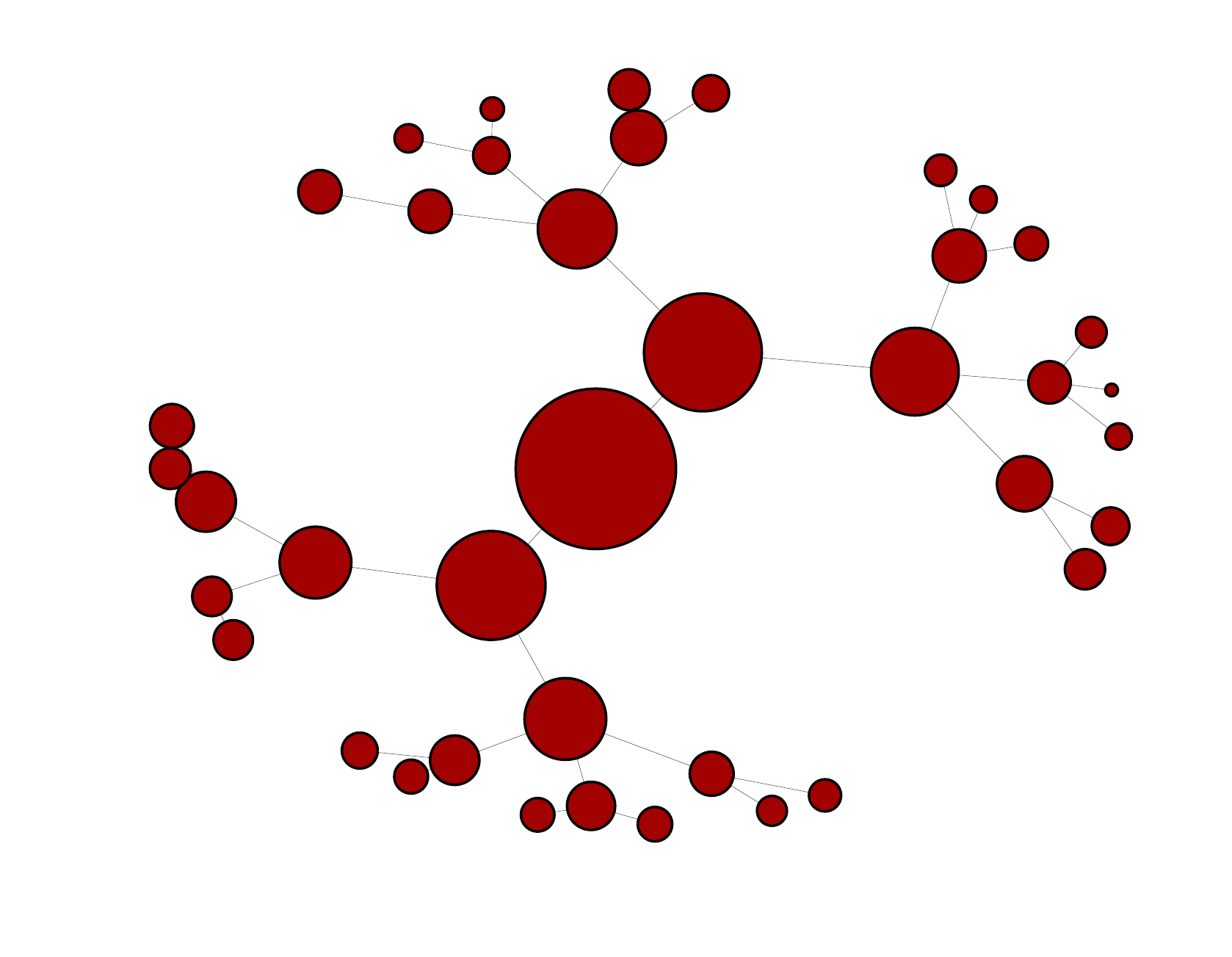}
\includegraphics*[width=.32\linewidth]{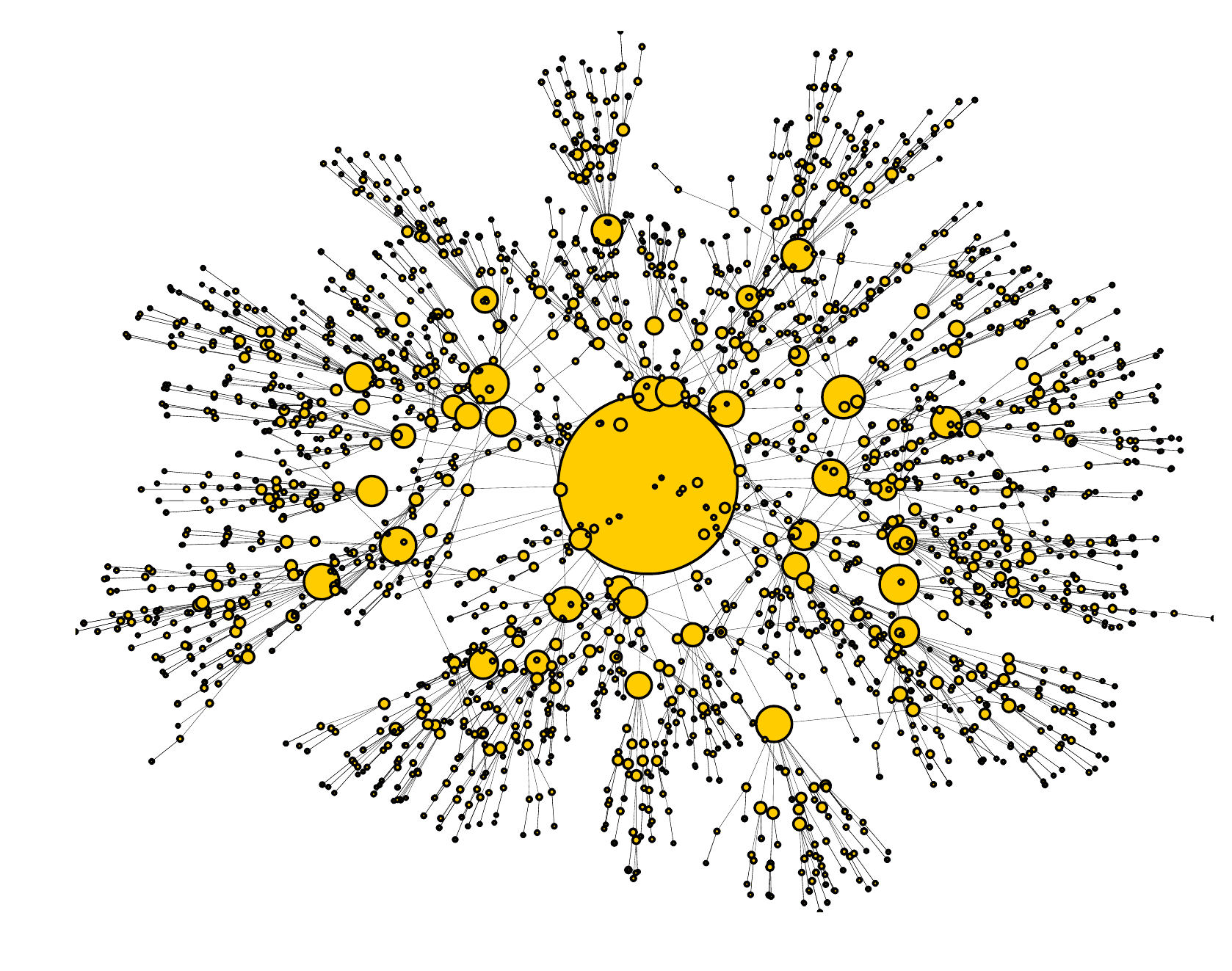}
\put(-475,110) {a)}
\put(-310,110) {b)}
\put(-155,110) {c)}
\caption{
(Color Online).
Hierarchical partition samples, or trees $\mathcal{T}$, computed from the Power-grid empirical network using: a) Infomap (left) , b) the HSBM method (middle) and c) RL (right).
The trees contain 1099, 40 and 2879 sub-communities, respectively.
The size of the sub-communities are proportional to the number of network nodes they contain.
The spring-layout is used to distribute the nodes on the plot~\cite{networkx}.
Clearly, the different community detection methods find significantly different hierarchies.
}
\label{fig:8}
\end{figure*}

\begin{figure*}
\includegraphics*[width=.24\linewidth]{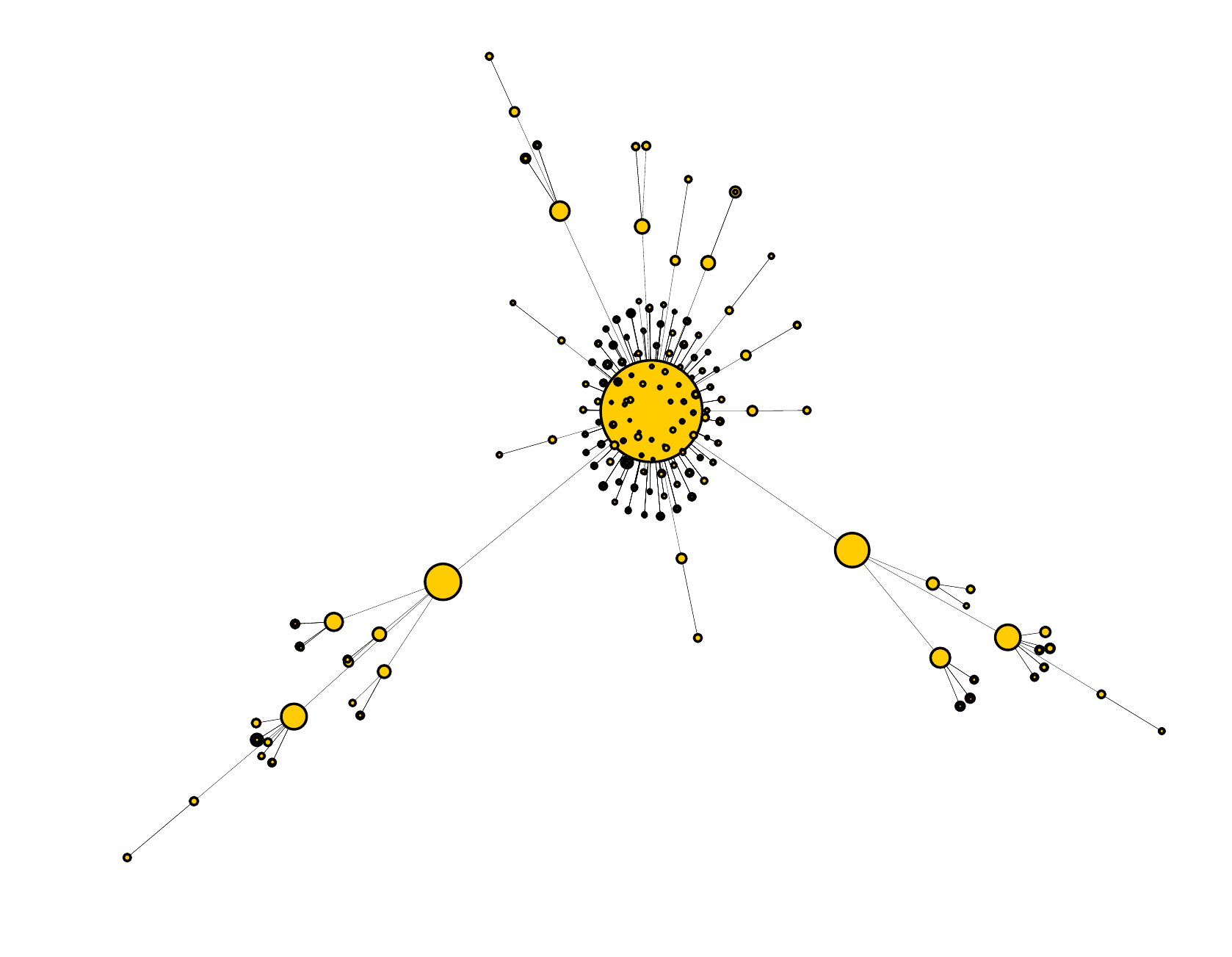}
\includegraphics*[width=.24\linewidth]{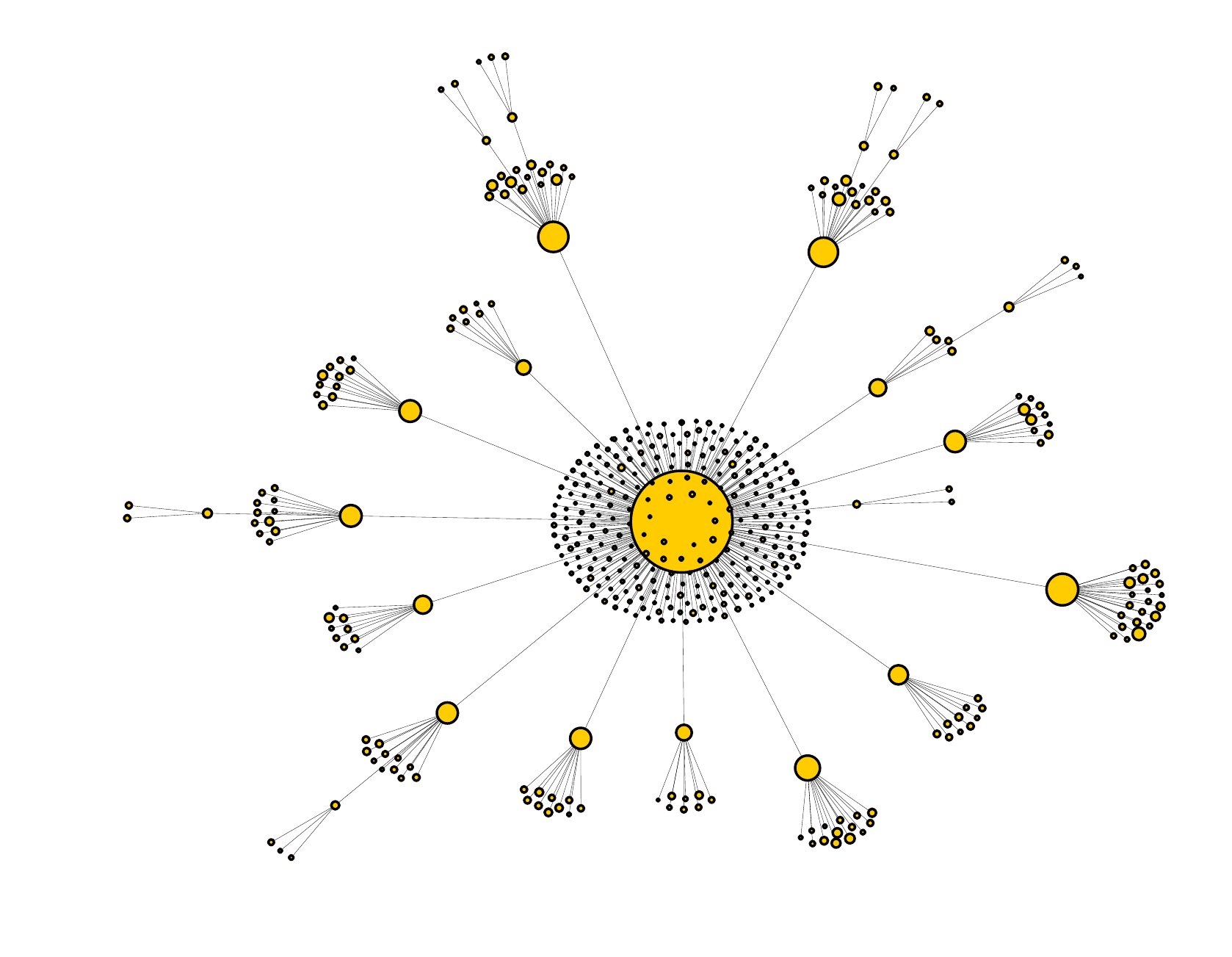}
\includegraphics*[width=.24\linewidth]{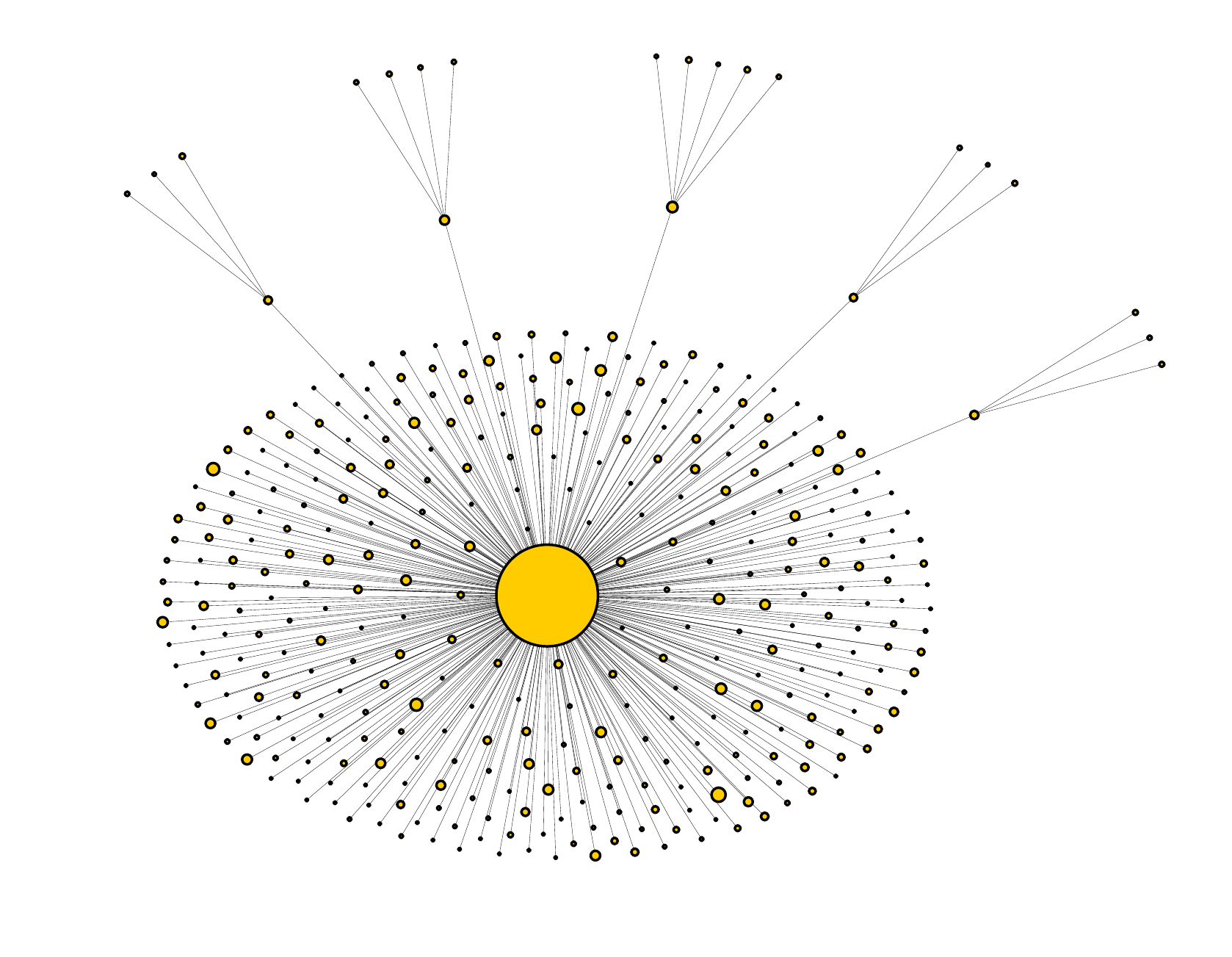}
\includegraphics*[width=.24\linewidth]{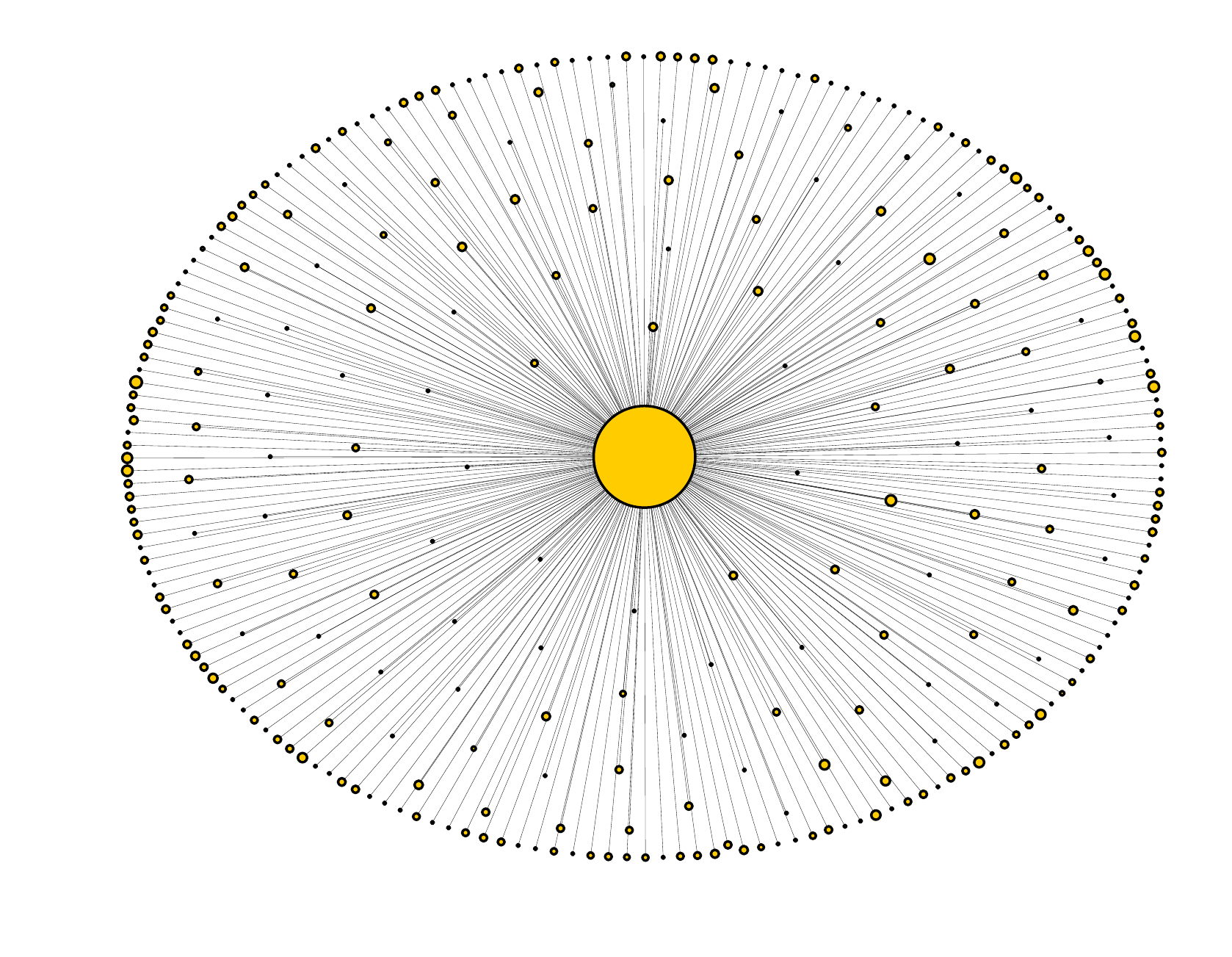}
\put(-470,90) {a)}
\put(-350,90) {b)}
\put(-225,90) {c)}
\put(-110,90) {d)}
\caption{
(Color Online).
Hierarchical partition samples $\mathcal{T}$, computed from the Network-science empirical network using Infomap.
Each panel correspond to a different level of link randomization: a) $f=0$, b) $f=0.2$, c) $f=0.5$ and d) $f=1$.
In Network-science network, the hierarchy is dominated by branches with almost no children at $f=0$, but two branches have considerable size and depth.
Then, as $f$ grows, the hierarchy evolves towards a simple  star, as shown in d).
}
\label{fig:10}
\end{figure*}

The  networks in Table~\ref{tab:1} (referenced therein) are the ones studied in the following analysis.
All of these networks have convenient characteristics: they are large enough to show relatively rich hierarchical community structures (e.g.~ Infomap returns up to five hierarchical levels for the case of the Power-grid~\cite{rosvall2011multilevel}), with diverse shape (e.g. compare the case of the Power-grid in Fig.~\ref{fig:8}a, with the case of the Network-science in Fig.~\ref{fig:10}a), and small enough to keep the computation time bounded.
Originally, some of these networks had link weights, or self-loops.
For the sake of simplicity, such attributes are removed from the networks.
As an illustration of how different are the hierarchical community structures identified by the different community detection methods, Fig.~\ref{fig:8} shows the results for the Power-grid network.
In this figure, it is apparent that the different methods provide substantially different results.

In order to enrich the analysis, the topology of the empirical networks is shuffled, following the same procedure applied to the Sierpinski networks (cf.~Section~\ref{sss:ahn}).
In this way, the obtained hierarchies are analyzed as a function of the fraction $f$  of randomized links.

\begin{figure*}
\includegraphics*[width=.32\linewidth]{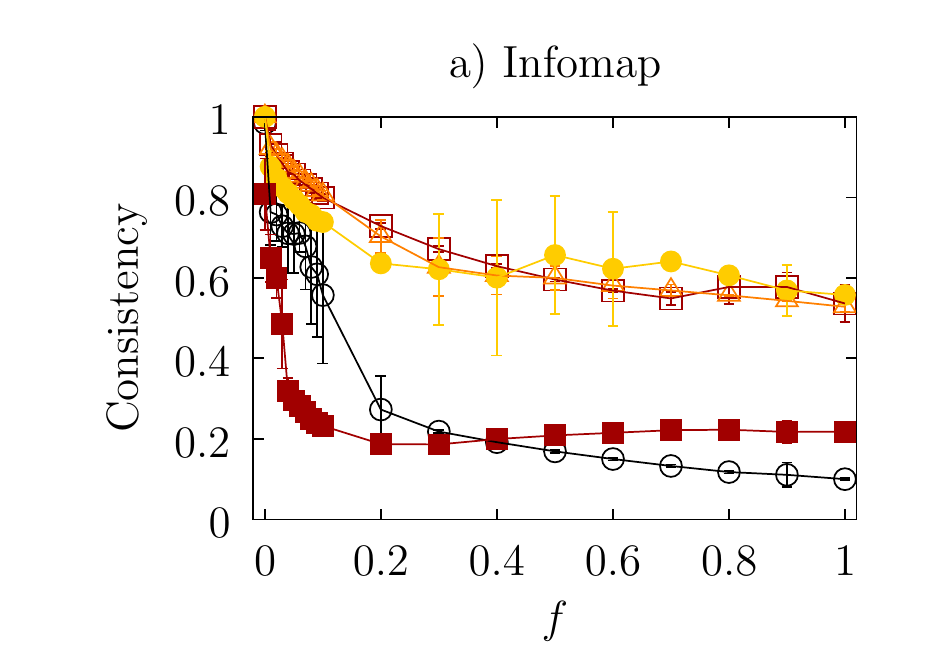}
\includegraphics*[width=.32\linewidth]{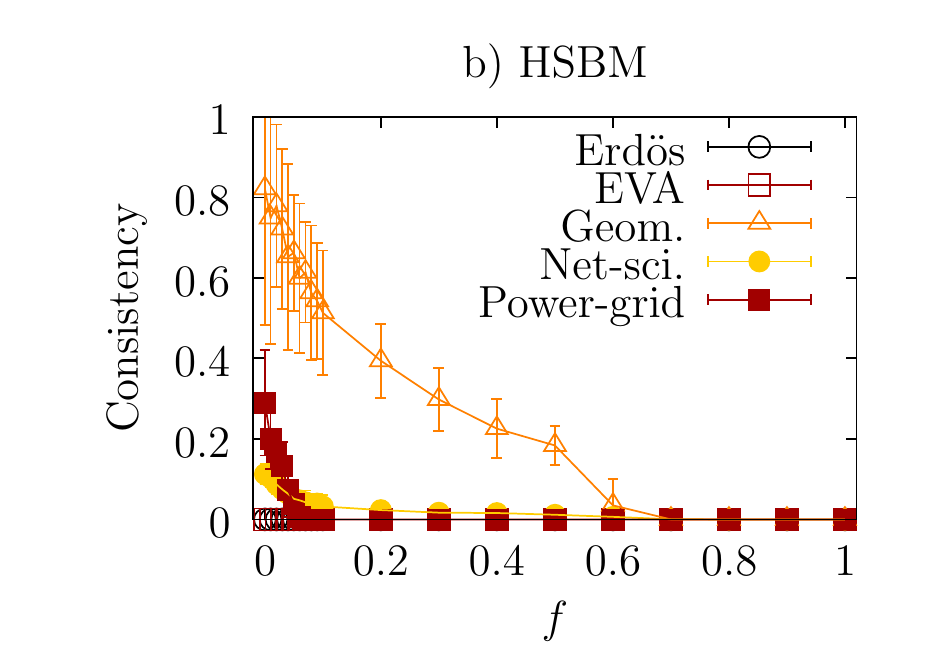}
\includegraphics*[width=.32\linewidth]{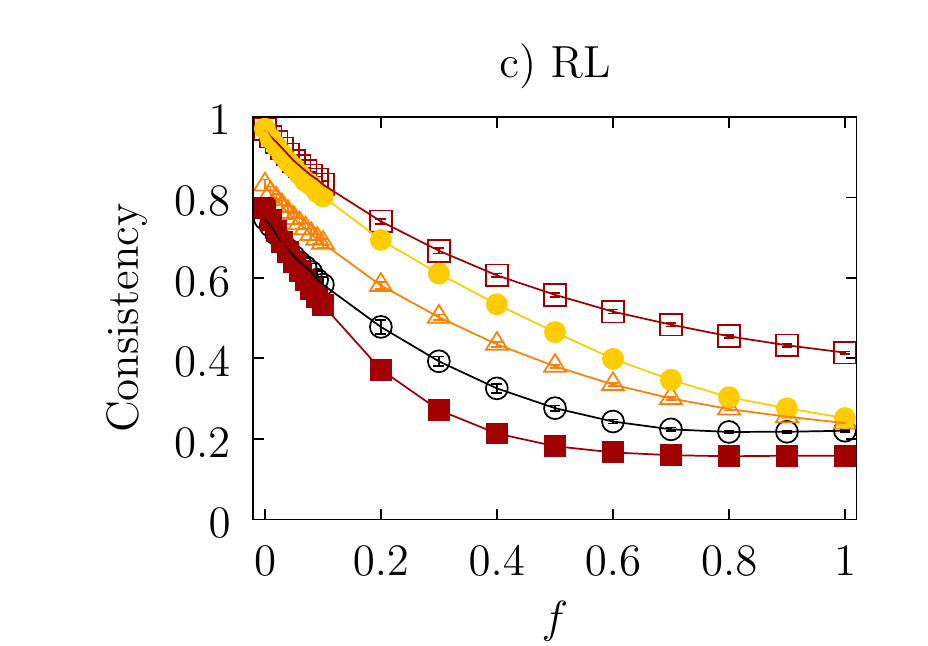}
\caption{
(Color Online).
The consistency is plotted for the different empirical networks in Table~\ref{tab:1}, as a function of the fraction of randomly rewired links, $f$, and for the different community detection methods: a) Infomap, b) HSBM, and c) RL.
In all cases, the bars represent standard-deviations around the mean.
}
\label{fig:9}
\end{figure*}

Firstly, Infomap is used to study the consistency of the empirical networks.
The results are shown in Fig.~\ref{fig:9}a.
For some networks, like the Power-grid and the Erd\H{o}s networks, the identified hierarchies are largely affected by the randomization procedure, i.e.~the consistency quickly decays with $f$.
This is particularly reasonable for the Power-grid network, as its hierarchy is embedded into space; 
reshuffling the links  attenuates the embedding, rapidly destroying its spatial nature~\cite{rozenfeld2010small,barthelemy2011spatial,popovic2012geometric}.
There exist other networks  like EVA, 
Geometry~and Network-science, 
for which the consistencies of the identified hierarchies seem quite robust to the randomization procedure.
This can be interpreted in two ways:
on the one hand, this suggests that the hierarchical community structure is mainly determined by the node degrees in the networks or some other topological property that is not destroyed by the randomization procedure.
On the other other hand, it may indicate that the the relatively large values of consistency are not significant from a hierarchical point of view.
A closer inspection to the Network-science~network reveals that the latter possibility is the cause.
Specifically, just a relatively small fraction of the network has a rich hierarchical structure with up to 4 levels. The rest of the network nodes are identified as communities at depth $l=1$, which have no children sub-communities (see Fig.~\ref{fig:10}a).
The hierarchical part is washed out as $f$ grows, eventually leading to a star-like structure (see Figs.~\ref{fig:10}b,~\ref{fig:10}c~and~\ref{fig:10}d).
The relatively large consistency values for large $f$ are the outcome of random coincidences occurring for these star-like structures.

Secondly, the HSBM method is used for the analysis and the results are shown in Fig.~\ref{fig:9}b.
Overall, a small consistency is obtained.
This is because the HSBM method often finds a single community, except for the 
Geometry network.
This suggests that the HSBM finds a rich hierarchical structure for the Geometry network in the form of nested block models.
However,  a closer inspection indicates that the HSBM identifies simply two large communities, i.e.~there is no hierarchy, similar to what is found for the Network-science network for the case of Infomap.
This explains the slow decay of the consistency curve.

Thirdly, the consistency is studied using the RL method.
The results are shown in Fig.~\ref{fig:9}c.
In all cases, the consistency presents a smooth decay as a function of the randomization $f$.
This is not a surprise because RL tends to return trees with a large number of sub-communities and levels.
Therefore, the small changes occurring for increasing $f$  lead to small changes in the consistencies.

\begin{figure*}
\includegraphics*[width=.32\linewidth]{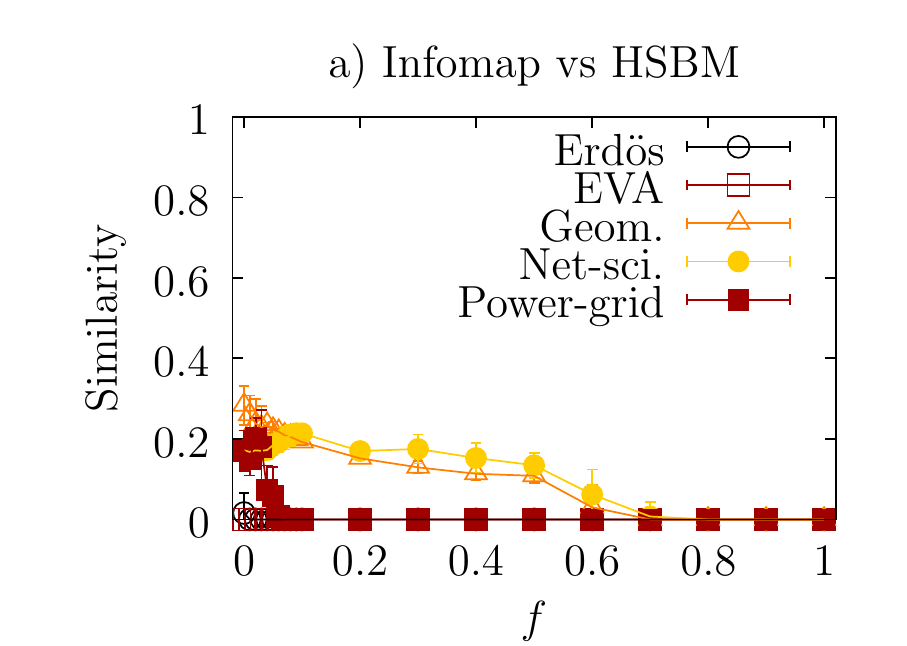}
\includegraphics*[width=.32\linewidth]{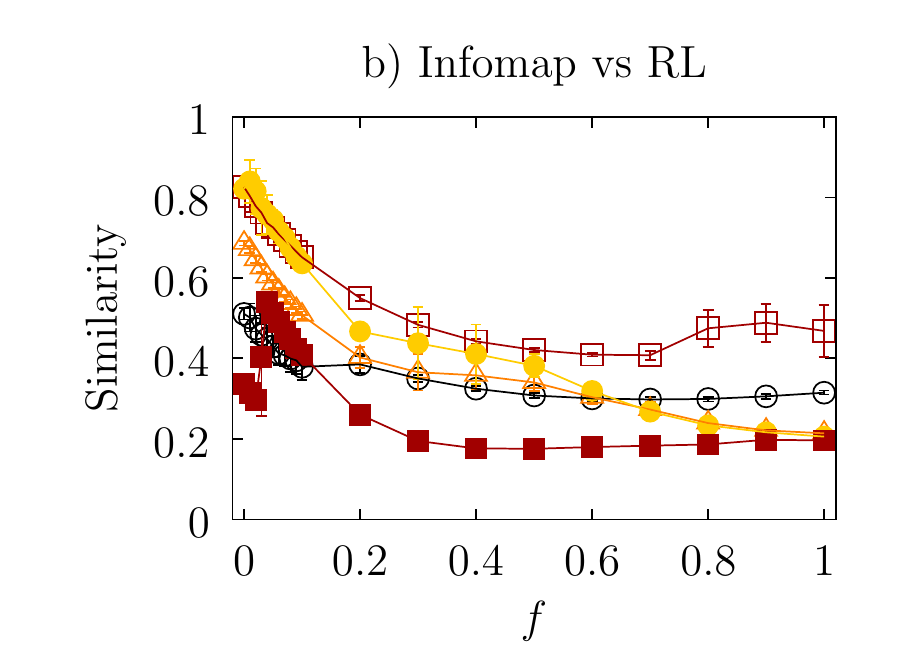}
\includegraphics*[width=.32\linewidth]{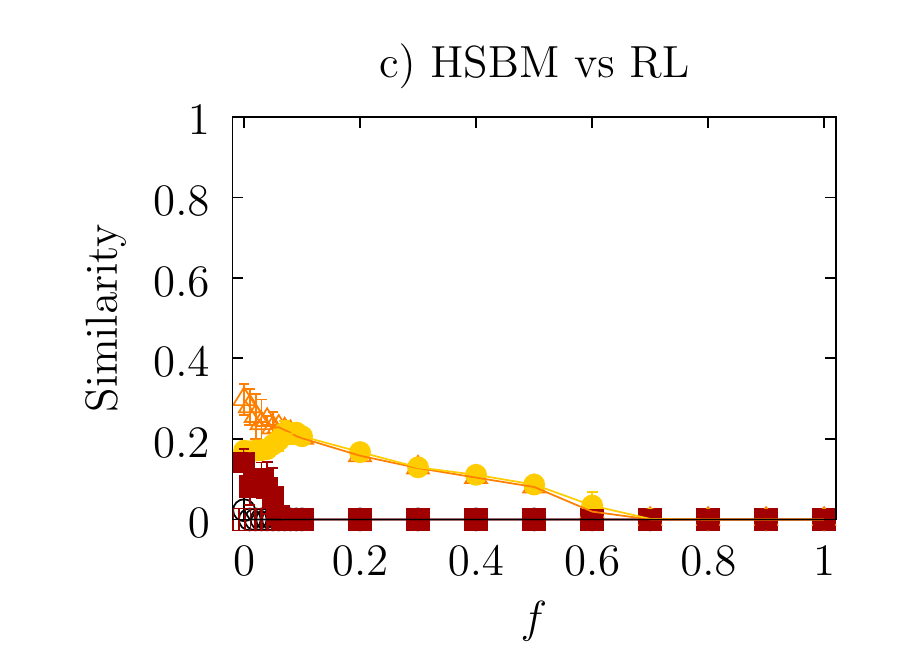}
\caption{
(Color Online).
The similarity is plotted for different empirical networks as a function of the fraction of randomly rewired links $f$, and the different pairs of community detection methods: a) Infomap vs HSBM, b) Infomap vs HSBM, and c) HSBM vs RL.
In all cases, the bars represent standard-deviations around the mean.
}
\label{fig:11}
\end{figure*}

In Figs.~\ref{fig:11}a~and~\ref{fig:11}c, the average similarity between the HSBM method and the other two methods is shown as a function of $f$.
Not surprisingly, the  values obtained are small.
However, it is interesting to note that, in certain cases, the similarity is larger than the corresponding values for the consistency. For example, cf.~the Network-science network in Fig.~\ref{fig:9}b and Fig.~\ref{fig:11}c, for small values of $f$.
Even though at a first glance this may seem contradictory, the explanation is simple.
The HSBM tends to return trivial hierarchies, yielding a value of zero for the hierarchical mutual information. Then, when the consistency is computed, the number of terms contributing with zero to the average $\avrg{i(\mathcal{T};\mathcal{T}')}$ 
is proportional to $1-p^2$, where $p$ is the probability for the HSBM to produce a non-trivial hierarchy.
On the other hand, such probability is $1-p$ for the case of the similarity because neither Infomap and nor RL produce trivial hierarchies.
In other words, the chances for zero terms to occur in the case of the consistency is significantly larger than for the case of the similarity.

In Fig.~\ref{fig:11}b, the similarity compares the results for Infomap and RL.
A sharp peak can be appreciated at $f\approx 0.05$.
This is because Infomap returns a sudden change over the number of identified hierarchies.
Namely, the hierarchies pass from having $\approx 4$ communities at level $l=1$, to up to $\approx 40$.
This large number of communities at level $l=1$ is always present for the RL.
Therefore, the sharp increase occurs when the number of communities at level $l=1$ becomes large for Infomap, i.e.~when it becomes similar for both methods.

\begin{table}
\caption{
Information summary about the empirical network datasets used in the calculations.
$N$ is number of nodes and $M$ number of links. 
Erd\H{o}s, Network-science and Geometry are scientific-collaboration networks.
The Power-grid is technological, and EVA is a network of corporate inter-relationships.
The networks marked with * were originally weighted.
}
\label{tab:1}
\begin{center}
\begin{tabular}{|c|c|c|c|}
\hline 
Network & $N$ & $M$ & Ref. \\ 
\hhline{|=|=|=|=|}
Power-grid & 4,941 & 6,594 &  \cite{watts1998collective} \\ 
\hline 
Erd\H{o}s & 6,927 & 11,850 & \cite{grossman2002erdos,pajek} \\
\hline 
Network-science* & 1,589 & 2,742 & \cite{newman2006finding,pajek} \\
\hline 
Geometry* & 7,343 & 11,898 & \cite{jones2002computational,pajek} \\
\hline
EVA & 8,497 & 7,970 & \cite{norlen2002eva,pajek} \\
\hline
\end{tabular} 
\end{center}
\end{table}

\subsubsection{Temporal Networks}

In this section, the subject of study is slightly modified.
Specifically, the study of traditional complex networks is replaced by the study of correlation matrices computed from the log-returns of stock prices in the S\&P500~\cite{mantegna1999hierarchical,bonanno2003correlation,tumminello2010correlation}.
The data is obtained from {\it Yahoo! Finance}~\cite{sp500_yahoofinance}.
In general, the correlation matrices can be considered as weighted dense networks.

Complex networks are not necessarily static, but change in time~\cite{holme2012temporal}.
The temporal aspect of a complex network could have dramatic consequences for the behavior of the associated system~\cite{starnini2012random,pfitzner2013betweenness,scholtes2014causality}.
The correlation matrices of the S\&P500 -- and the associated hierarchical community structures -- can be studied in their time evolution~\cite{bazzi2014community,macmahon2015community,granell2015benchmark}. Therefore, we use the hierarchical mutual information to investigate the evolution of the hierarchical community structure of the financial activity in the S\&P500.

The data encompasses the 390 stocks which  uninterruptedly cover the 3522 working days from January $1^{\mathrm{st}}$, 1998 until December $31^{\mathrm{st}}$, 2011, according to Yahoo! Finance.
Each matrix entry of the correlation matrices is given by 
\begin{equation}
\label{eq:9}
C_{ss'}=\frac{\mathrm{Cov}(X_s,X_{s'})}{\sqrt{\mathrm{Var}(X_s)\mathrm{Var}(X_{s'})}}.
\end{equation}
Specifically, the r.h.s.~of Eq.~\ref{eq:9} is the {\it cross-correlation} between the time series $X_s(t)$ and $X_{s'}(t)$, corresponding to the stocks $s$ and $s'$, respectively.
In general, cross correlation matrices have off-diagonal entries in $[-1,1]$, while diagonal entries are equal to one.
To simplify the analysis, the correlation matrices are transformed according to the expression~\cite{heimo2008detecting},
$w_{ss'}=|C_{ss'}|-\delta_{ss'}$.
The transformation returns a weighted network of non-negative entries and zero diagonals.
The transformed networks are the ones used for the computation of the hierarchies.
Even though more sophisticated approaches  exist (see for example Ref.~\cite{macmahon2015community}),
for the sake on simplicity the approach taken is the one described above.

To perform a temporal analysis, different correlation matrices, or weighted networks, are computed by processing the data over different time windows $[t,t+T]$, where $t$ is the initial day of the time window, and $T$ the window duration, measured in days.

In the following analysis, only the RL community detection method is used, this is because the other two methods typically return trivial communities.
More specifically, the other two methods fail to find communities because the correlation networks are dense~\cite{kawamoto2015estimating}.
On the other hand, RL is more sensitive to small link-weights differences, and therefore, it is able to find communities in the dense matrices, but at risk of over-fitting (see section~\ref{ss:art_fid}).
As it was already mentioned, more sophisticate methods can be used to mitigate these undesired tendencies~(see section~\ref{sss:cdm}).
However, such experiments are left for future works.

\begin{figure}
\includegraphics*[width=.9\linewidth]{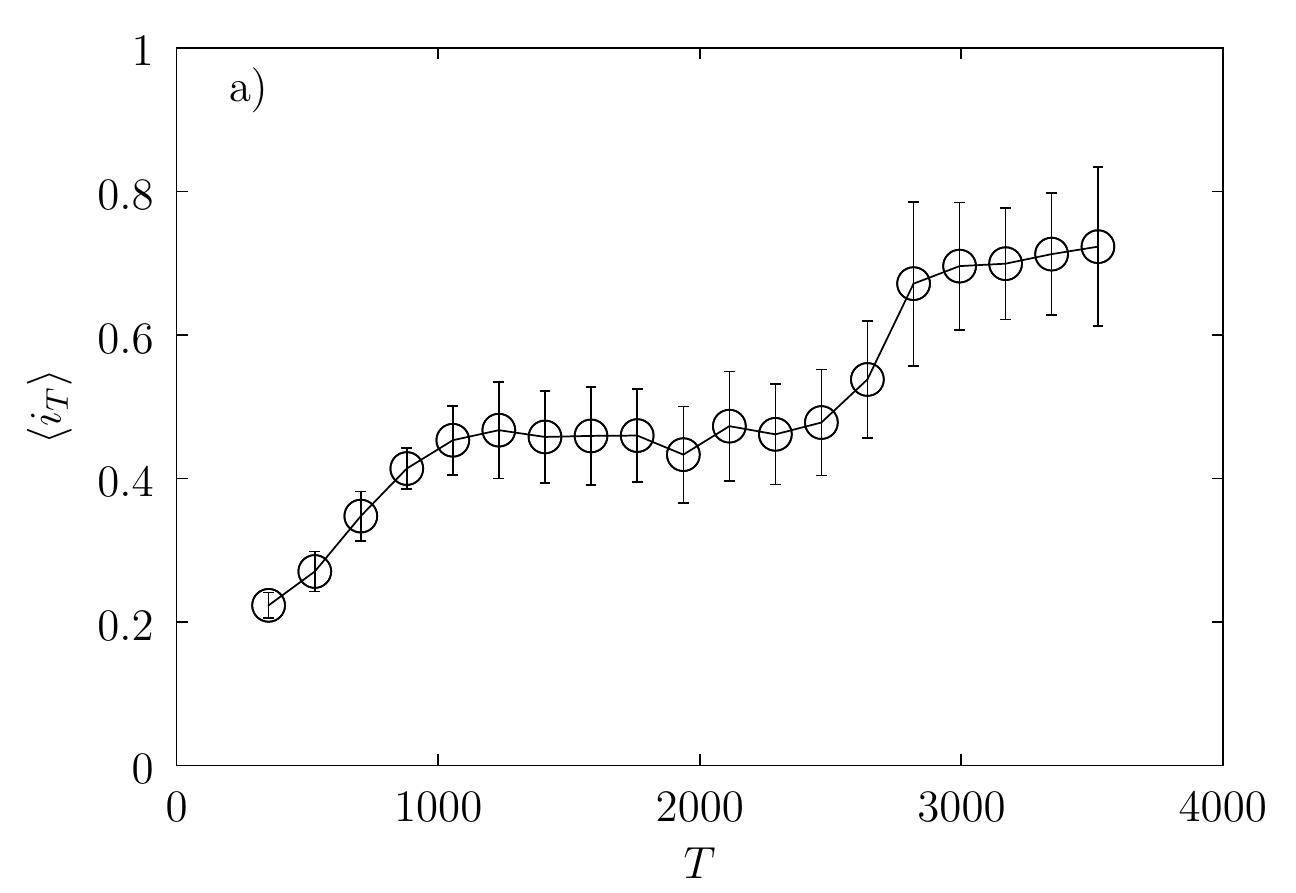}
\\
\includegraphics*[trim=35 0 0 0 ,width=.9\linewidth]{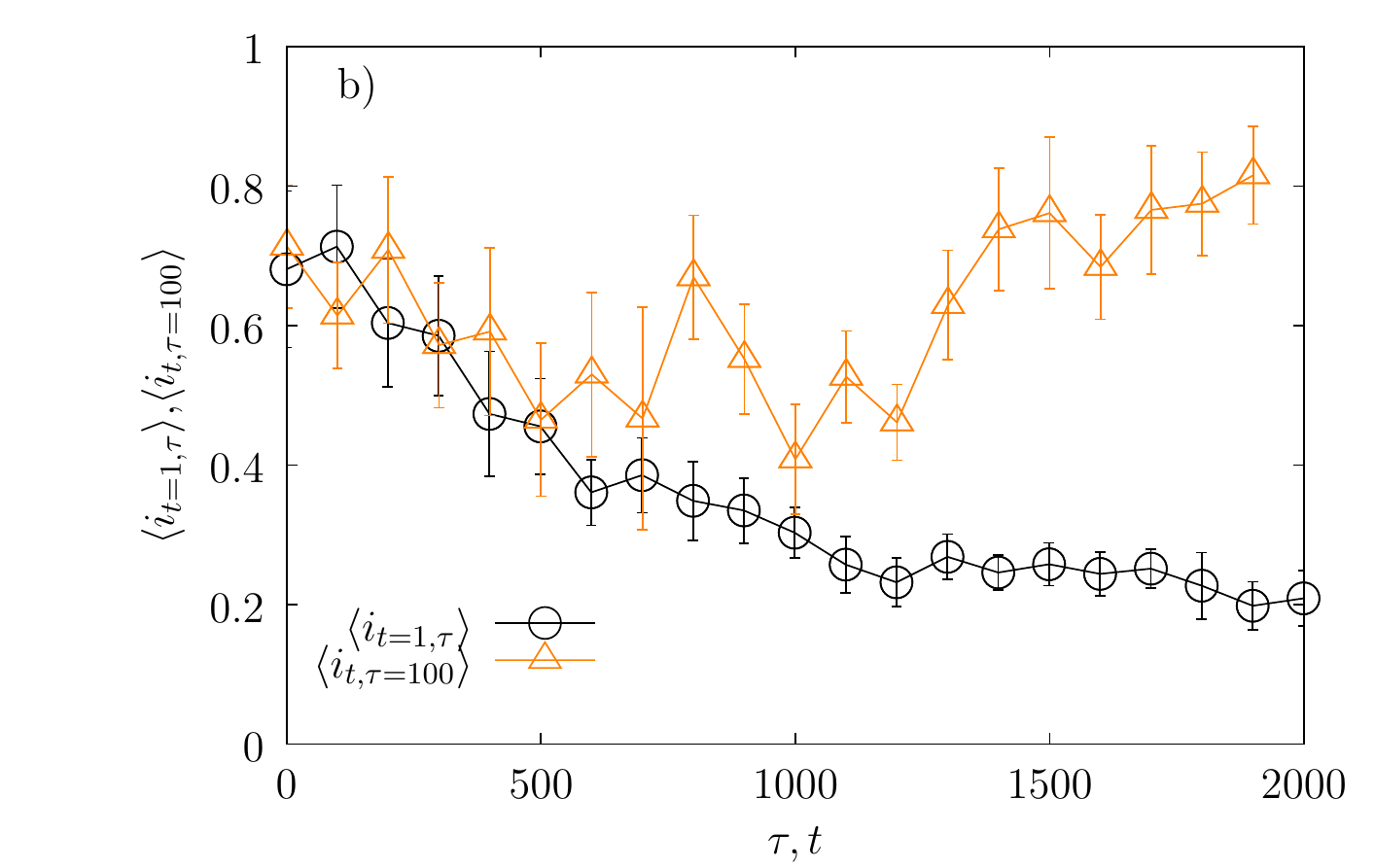}
\caption{
(Color Online).
Temporal analysis of the hierarchical community structure of correlation matrices.
The matrices are computed from log-returns of the time series of stock prices in the S\&P500.
In a), the scale-similarity $\avrg{i_T}$, determines how the hierarchies change with the time length $T$ of the  time window over which the data is processed.
In b), two different comparisons are presented using the auto-similarity $\avrg{i_{t,\tau}}$.
With circles, $\avrg{i_{t=1,\tau}}$ determines how similar are the hierarchies at day one, with the hierarchies $\tau$ days after.
With triangles, $\avrg{i_{t,\tau=100}}$ determines how similar are the hierarchies of consecutive time windows, separated by 100 days, as time $t$ evolves.
In all cases, the bars represent standard-deviations around the mean.
}
\label{fig:12}
\end{figure}

Two sets of experiments are analyzed; in both cases, for each computed weighted network, 50 hierarchical community structures, or trees, are computed.
In the first set of experiments, we analyze how the integration time, or time windows length $T$, affects the detected hierarchies.
For this purpose, we compute the following average normalized hierarchical mutual information, 
$$
\avrg{i_T}:=\avrg{i(\mathcal{T}_{T_{\mathrm{max}}};\mathcal{T}_T)}.
$$
We call this quantity, the {\it temporal-scale hierarchical similarity}, or simply, the {\it scale-similarity}.
It compares hierarchies obtained from the full-length time window, against hierarchies obtained from time windows of length $T$.
In all cases, the initial time is chosen to be the first day, $t=0$.
In Fig.~\ref{fig:12}a, $\avrg{i_T}$ is plotted as a function of $T$.
As it can be seen, the larger is $T$, the larger is $\avrg{i_T}$. In other words, the expected behavior is observed because, the larger is $T$ the more similar $\mathcal{T}_T$ and $\mathcal{T}_{T_{\mathrm{max}}}$ become in average.
In particular, a {\it plateau} exists for $1000 \lesssim T \lesssim 3000$.
This last observation suggests that changes do not occur smoothly, but different hierarchical structural properties emerge at different time scales.

In the second set of experiments, $T$ is fixed at $1500$ days and trees are computed out of networks corresponding to different regions in the time line.
More specifically, we introduce the {\it temporal hierarchical auto-similarity} -- or {\it auto-similarity} -- which is defined as
$$
\avrg{i_{t,\tau}}:=\avrg{i(\mathcal{T}_{t};\mathcal{T}_{t+\tau})}.
$$
The auto-similarity compares two set of hierarchies.
The first set is computed from the data in the time window $[t,t+T]$, and the other set from the time window defined $\tau$ days after.
We analyze the auto-similarity varying $\tau$ for fixed $t=1$, and varying $t$ for fixed $\tau=100$.
In the first case, we study how the time separation $\tau$ affects the hierarchy, and in the second case we compare hierarchies corresponding to consecutive time windows as time evolves.
In Fig.~\ref{fig:12}b, both quantities are plotted.
On the one hand, the auto-similarity $\avrg{i_{t=1,\tau}}$ decays as the time separation $\tau$ grows (circles), i.e. the hierarchy drift away from the initial structure.
On the other hand, the auto-similarity fluctuates around $\avrg{i_{t,\tau=100}}\approx 0.7$ (triangles), indicating that the hierarchies of consecutive time windows always share a significant amount of information.

\section{Discussion, conclusions and future work}
\label{sec:conclu}

In this work, the hierarchical mutual information has been introduced, a tool that generalizes the standard mutual information for the comparison of hierarchies.
More specifically, for the comparison of hierarchical partitions, which take the form of trees where parts are subsequently subdivided further into sub-parts and so on.
The hierarchical mutual information can be used to compare the hierarchical community structure of complex networks, in analogous way as the standard mutual information can be used to compare standard community structures.

We define here a normalized hierarchical mutual information.
The traditional normalized mutual information satisfy certain properties; it is a quantity lying in $[0,1]$, and is equal to one if and only if the compared partitions are exactly equal.
If the normalized hierarchical mutual information behaves correctly, it should satisfy analogous properties.
The appropriate behavior of the normalized hierarchical mutual information is extensively tested in numerical experiments.
The test include artificially generated hierarchical partitions, and the hierarchical community structure of artificially and empirical complex networks.
In all the experiments, the normalized hierarchical mutual information is found to behave correctly.
However, it should be mentioned that a formal proof of the correct behavior is not provided in the present work.

The experiments also illustrate the overall behavior of the hierarchical mutual information.
On the one hand, when comparing artificially generated hierarchies against correspondingly randomized ones, the normalized hierarchical mutual information was found to decrease with the level of randomization.
On the other hand, a level by level randomization analysis of the hierarchies indicated that, 
the larger the number of randomized levels, the faster the normalized hierarchical mutual information decays with the randomization.
Another interesting finding was that the normalized hierarchical mutual information never decays to zero.
This effect, also present in the standard normalized mutual information, occurs because random (hierarchical) partitions in finite systems share information just by chance.

The experiments also constitute examples of how the hierarchical mutual information can be used to analyze the hierarchical community structure of complex networks.
Specifically, the hierarchical community structure of artificial and empirical networks were studied.
In the analysis, different popular community detection methods were utilized, and the results compared.
The results were tested on two network models and five empirical networks.
It was found that the different methods can return significantly different hierarchical community structures.
The normalized hierarchical mutual information correctly identifies these differences.
It was also shown that the normalized hierarchical mutual information can be used to compare the detected hierarchies against the natural, reference ones in the different network models.
In particular, when the parameters of the network models are appropriate, and the network models tend to generate networks with the expected hierarchical structures, the normalized mutual information between the identified hierarchies and the expected ones tends to grow.

In another set of experiments, the normalized hierarchical mutual information was used to compare the hierarchical community structure of the different networks -- the networks generated by the models, and the empirical networks -- against that 
of correspondingly randomized networks.
As expected, the normalized mutual information was found to decay with the level of randomization.
In a final example, the time evolution of the hierarchical community structure of correlation matrices was analyzed.
Specifically, we considered correlation matrices computed from the log returns of stock prices in the S\&P500.
This final example epitomizes how the hierarchical mutual information is useful to study the evolution of temporal networks.
In the analysis, the normalized hierarchical mutual information showed that the hierarchical community structure of the correlations of stocks slowly changes in time, but exhibiting  important changes at different  times-scales.

The present work opens several possibilities for future research.
The mathematical framework behind the hierarchical mutual information can be used to generalize other information measures, like generalizing the variation of information~\cite{meilua2007comparing}.
On a different line of research, the normalized hierarchical mutual information can be used to systematically benchmark, and compare, the different community detection methods in existence.
Another interesting future line of research concerns the comparison of phylogenetic trees~\cite{robinson1981comparison,dasgupta1997distances,shi2013distances,van2014approximation}, where the hierarchical mutual information could have useful applications.
Finally, the normalized hierarchical mutual information can be used to compare the identified hierarchies against corresponding {\it ground-truth hierarchies} that different data sets might have available.
The above examples go without mentioning the ample possibilities of using and extending this methodology in the many fields where hierarchical communities structures are identified.

\section{Acknowledgments}
J.I.P and G.C. acknowledge support from: FET IP Project MULTIPLEX nr. 317532. FET Project SIMPOL nr. 610704, FET project DOLFINS nr. 640772.
CJT acknowledges financial support from the URPP on Social Networks, Universit\"at Z\"urich, Switzerland.
We also acknowledge useful comments by A. Clauset, M. Rosvall, T. Peixoto and F. Queyroi. 



%


\end{document}